%% file: main.tex
\documentclass[
sigconf, 
nonacm, 
authorversion,
natbib,  
balance  
]{acmart}

\usepackage{amsmath}
\usepackage{amsfonts}
\usepackage[utf8]{inputenc}
\usepackage{array}
\usepackage{stmaryrd}
\usepackage{tikz}
\usepackage{comment}
\usepackage{xspace}
\usepackage{listofitems}
\usepackage{graphicx}
\usepackage[final]{listings}
\usepackage{hyperref}
\usepackage{enumitem}
\usepackage{amsthm}
\usepackage{titlesec}

\newif\ifstreamexamples
\streamexamplestrue
\newif\ifzsetexamples
\zsetexamplestrue
\newtheoremstyle{note} 
{2pt} 
{2pt} 
{}    
{}    
{\bfseries} 
{:}   
{.5em}
{}    

\numberwithin{equation}{section}

\graphicspath{ {.} }
\lstdefinelanguage{ddlog}{
  language=Java, 
  morekeywords={input, output, typedef, relation, typedef, bool, not,
    string, bit, extern, function, var, for, match, skip, in, integer, 
    Aggregate, FlatMap},
  deletestring=[b]{'}
}
\hypersetup{
  colorlinks   = true,    
  urlcolor     = blue,    
  linkcolor    = blue,    
  citecolor    = red      
}
\hypersetup{final}

\usetikzlibrary{shapes, arrows.meta, positioning}
\tikzstyle{block}=[draw,fill=white,rectangle]
\tikzstyle{every node}=[font=\small]

\theoremstyle{note}
\newtheorem{theorem}{Theorem}[section]
\newtheorem{lemma}[theorem]{Lemma}
\newtheorem{corollary}[theorem]{Corollary}
\newtheorem{definition}[theorem]{Definition}
\newtheorem{proposition}[theorem]{Proposition}

\newtheorem{algorithm}[theorem]{Algorithm}
\newcommand{\dbsp}{DBSP\xspace}

\newcommand{\defined}[1]{\textbf{#1}\index{}}
\newcommand{\zr}{$\Z$-set\xspace}
\newcommand{\zrs}{$\Z$-sets\xspace} 

\newcommand{\code}[1]{\mbox{\texttt{#1}}}
\newcommand{\Z}{\mathbb{Z}}  
\newcommand{\N}{\mathbb{N}}  
\newcommand{\B}{\mathbb{B}}  
\newcommand{\R}{\mathbb{R}}  
\newcommand{\stream}[1]{\ensuremath{\mathcal{S}_{#1}}}
\newcommand{\streamf}[1]{\ensuremath{\overline{\mathcal{S}_{#1}}}}
\newcommand{\zm}{\ensuremath{z^{-1}}} 
\newcommand{\I}{\mathcal{I}}  
\newcommand{\D}{\mathcal{D}}  
\newcommand{\inc}[1]{{#1}^{\Delta}}
\newcommand{\distinct}{\mathit{distinct}}  
\newcommand{\secref}[1]{\S\ref{#1}}  
\newcommand{\refsec}[1]{\secref{#1}}

\newcommand{\id}{\ensuremath{\mathit{id}}} 
\newcommand{\isset}{\mbox{isset}}
\newcommand{\ispositive}{\mbox{ispositive}}
\newcommand{\defn}{\stackrel{\textrm{\scriptsize def}}{=}}
\newcommand{\map}{\mbox{map}}
\newcommand{\fix}[2]{\mbox{fix}\,#1.#2}
\newcommand{\lift}[1]{{\uparrow}#1}              
\newcommand{\norm}[1]{\| #1 \|} 
\newcommand{\zpp}[1]{\mbox{zpp}(#1)}
\newcommand{\makeset}{\ensuremath{\mbox{makeset}}}
\newcommand{\sv}[1]{ 
\setsepchar{ }
\readlist\arg{#1}
{[}
\begin{array}{cccccc}
    \arg[1] & \arg[2] & \arg[3] & \arg[4] & \arg[5] & \cdots
\end{array}
{]}
}

\newcommand{\scream}[2]{{\color{red} \textbf{#1}: #2}}
\newcommand{\val}[1]{\scream{VAL}{#1}}

\title{\dbsp: Automatic Incremental View Maintenance for Rich Query Languages}
\author{Mihai Budiu}
\affiliation{VMware Research}
\email{mbudiu@vmware.com}
\author
{Frank McSherry}
\affiliation{Materialize Inc.}
\email{mcsherry@materialize.com}
\author
{Leonid Ryzhyk}
\affiliation{VMware Research}
\email{lryzhyk@vmware.com}
\author
{Val Tannen}
\affiliation{University of Pennsylvania}
\email{val@seas.upenn.edu}

\begin{abstract}
Incremental view maintenance has been for a long time a central problem in database theory~\cite{gupta-idb93}.
Many solutions have been proposed for restricted classes of database languages,
such as the relational algebra, or Datalog.  These techniques do not naturally generalize to
richer languages.  In this paper we give a general
solution to this problem in 3 steps: (1) we describe a simple but expressive language
called \dbsp for describing computations over data streams; (2) we give a general algorithm for 
solving the incremental view maintenance problem for arbitrary \dbsp programs, and (3) we show
how to model many rich database query languages (including the full relational queries, 
grouping and aggregation, monotonic and non-monotonic recursion, and streaming aggregation) using \dbsp.  
As a consequence, we obtain efficient
incremental view maintenance techniques for all these rich languages.
\end{abstract}

\begin{document}

\maketitle

\input{intro}
\input{streams}
\input{relational}
\input{recursion}
\input{nested}

\input{extensions}
\input{implementation}
\input{related}
\input{conclusions}

\bibliographystyle{plainurl}
\bibliography{main}

\appendix
\input{extra}

\end{document}

%% file: intro.tex
\section{Introduction}\label{sec:ntro}

In this paper we present a simple mathematical theory for modeling
streaming and incremental computations.  This model has immediate
practical applications in the design and implementation of streaming databases
and incremental view maintenance.  Our model is based on mathematical formalisms 
used in discrete digital signal processing (DSP)~\cite{rabiner-book75},
but we apply it to database computations.  
Thus, we have called it ``\dbsp''.  \dbsp is inspired from Differential 
Dataflow~\cite{mcsherry-cidr13} (DD), and started as an attempt to provide a simpler
formalization of DD than the one of Abadi et al.~\cite{abadi-fossacs15} 
(as discussed in \secref{sec:related}), but has evolved behind that purpose.

The core concept of \dbsp is the \emph{stream}: a stream $s$ with type 
$\stream{A}$ maps ``time'' moments $t\in\N$ 
to values $s[t]$ of type $A$; think of it as an "infinite vector".
A streaming computation is a function that
consumes one or more streams and produces another stream.  We depict
streaming computations with typical DSP box-and-arrow diagrams (also called ``circuits''),
where boxes are computations and streams are arrows, as in the following diagram,
which shows a stream operator $T$ consuming two input streams $s_0$ and $s_1$ 
and producing one output stream $s$:

\begin{center}
\begin{tikzpicture}[auto,>=latex,minimum width=.5cm]
  \node[] (input0) {$s_0$};
  \node[below of=input0,node distance=.3cm] (dummy) {};
  \node[below of=dummy,node distance=.3cm] (input1) {$s_1$};
  \node[block, right of=dummy] (T) {$T$};
  \node[right of=T] (output) {$s$};
  \draw[->] (input0) -- (T);
  \draw[->] (input1) -- (T);
  \draw[->] (T) -- (output);
\end{tikzpicture}
\vspace{-.2cm}
\end{center}

We generally think of streams as sequences of \emph{small} values,
and we will use them in this way.
However, we make a leap of imagination and also treat a \emph{whole database} as a stream value.
What is a stream of databases?  It is a \emph{sequence of database
snapshots}.  We model the time-evolution of a database $DB$ as a
stream $DB \in \stream{SCH}$, where $SCH$ is the database schema.
Time is not the wall-clock time, but essentially a counter
of the \emph{sequence of transactions} applied to the database. 
Since transactions are linearizable,
they have a total order, which defines a linear time $t$ dimension:
the value of the stream $DB[t]$ is the snapshot of the 
database contents after $t$ transactions have been applied.  We assume that $DB[0] = 0$, i.e., the database starts empty.

Database transactions also form a stream $T$, a stream of \emph{changes},
or \emph{deltas} that are applied to our database.   
The database snapshot at time $t$ is the cumulative result of applying all 
transactions in the sequence up to $t$: $DB[t] = \sum_{i \leq t} T[i] \defn \I(T)[t]$ (we make the notion of ``addition'' precise later.).
The operation of adding up all changes is \emph{stream integration}.
The following diagram expresses this relationship using the $\I$ operator for 
stream integration:

\begin{center}
\begin{tikzpicture}[auto,>=latex,minimum width=.5cm]
  \node[] (input) {$T$};
  \node[block, right of=input] (I) {$\I$};
  \node[right of=I] (output) {$DB$};
  \draw[->] (input) -- (I);
  \draw[->] (I) -- (output);
\end{tikzpicture}
\end{center}

Conversely, we can say that transactions are the \emph{changes} of a database, and write
$T = \D(DB)$, or $T[t] = DB[t] - DB[t-1]$.  This is the definition of stream differentiation, denoted by $\D$; this 
operation computes the changes of a stream, and is the 
inverse of stream integration.  \secref{sec:streams}
precisely defines streams, integration and differentiation, and analyzes
their properties. 

Let us apply these concepts to view maintenance.
Consider a database $DB$ and a query $Q$ defining a view $V$ as a function
of a database snapshot $V = Q(DB)$.  Corresponding to the stream of database snapshots
$DB$ we have a \emph{stream of view snapshots}: $V[t]$ is the
view's contents after the $t$-th transaction has been applied.  We show this
relationship using the following diagram:

\begin{center}
\begin{tikzpicture}[auto,>=latex,minimum width=.5cm]
  \node[] (input) {$DB$};
  \node[block, right of=input] (I) {$\lift{Q}$};
  \node[right of=I] (output) {$V$};
  \draw[->] (input) -- (I);
  \draw[->] (I) -- (output);
\end{tikzpicture}
\end{center}

The symbol $\lift{Q}$ (the ``lifting'' of $Q$) shows that the query $Q$ is applied
independently to every element of the stream of database snapshots $DB$.  
$\lift{Q}$ is a ``streaming query'' since it operates on a stream of values.
The incremental view maintenance problem requires an algorithm to  
compute the stream $\Delta V$ of \emph{changes} of the view $V$, i.e., $\D(V)$,
as a function of the stream $T$.  
By chaining these definitions together we get the following \defined{fundamental equation}
of the view maintenance problem: $\Delta V = \D(\lift{Q}(DB)) = \D(\lift{Q}(\I(T)))$,
graphically shown as:

\begin{center}
\begin{tikzpicture}[auto,>=latex,minimum width=.5cm]
  \node[] (input) {$T$};
  \node[block, right of=input] (I) {$\I$};
  \node[block, right of=I, node distance=1.5cm] (Q) {$\lift{Q}$};
  \node[block, right of=Q] (D) {$\D$};
  \node[right of=D] (output) {$\Delta V$};
  \draw[->] (input) -- (I);
  \draw[->] (I) -- node (db) {$DB$} (Q);
  \draw[->] (Q) -- (D);
  \draw[->] (D) -- (output);
\end{tikzpicture}
\end{center}

This definition can be generalized to more general streaming queries
$S: \stream{A} \to \stream{B}$ that are richer than lifted pointwise queries $Q$.
The incremental version of streaming query $S$ is denoted by $\inc{S}$ and is defined
according to the above equation, which can also be written as: $\inc{S} = \D \circ S \circ \I$. 

It is generally assumed that the changes to a dataset are much smaller than
the dataset itself; thus, computing on streams of changes may
produce significant performance benefits.

Applying the query \emph{incrementalization} operator $S \mapsto \inc{S}$ constructs
a query that computes directly on changes; however, the resulting query is no more
efficient than a query that computes on the entire dataset, because it uses an
integration operator to reconstitute the full dataset.
\secref{sec:incremental} shows how algebraic properties of the $\inc{\cdot}$ operator 
are used to optimize the implementation of $\inc{S}$:

\begin{enumerate}[nosep, leftmargin=0pt, itemindent=0.5cm, label=\textbf{(\arabic*)}]
\item The first property is that many classes of primitive operations have very efficient incremental
versions.  In particular, linear queries have the property $Q = \inc{Q}$.  Almost all
relational and Datalog queries are based on linear operators.  Thus, the incremental version
of such queries can be computed in time proportional to the size of the changes.  Bilinear 
operators (such as joins) have a more complex implementation, which nevertheless still performs 
work proportional to the size of the changes, but require storing an amount of data proportional
to the size of the relations.

\item The second key property is the chain rule:
$\inc{(S_1 \circ S_2)} = \inc{S_1} \circ \inc{S_2}$.  This rule gives the incremental
version of a complex query as a composition of incremental versions of its components.
It follows that we can implement any incremental query as a composition of primitive
incremental queries, \emph{all of which perform work proportional to the size of the changes}.
\end{enumerate}

Armed with this general theory of incremental computation, in \secref{sec:relational}  
we show how to model relational queries in \dbsp.  This immediately gives
us a general algorithm to compute the incremental version of any relational query.
These results are well-known, but they are cleanly modeled by \dbsp.

Applying DBSP to recursive queries requires extending this computational model.  
In \secref{sec:recursion} we introduce two additional operators: $\delta_0$
creates a stream from a scalar value, and $\int$ creates a scalar value from a stream.  
These operators can be used to implement computations with \code{while} loops.
So, in addition to modelling changing inputs and database, we also
use streams as a model for sequences of \emph{consecutive values of loop
iteration variables}.  With this addition \dbsp becomes rich enough to implement 
recursive queries.  \secref{sec:datalog} shows how stratified recursive
Datalog programs with negation can be implemented in \dbsp.  

In \secref{sec:nested} we use \dbsp to model computations on nested streams, where each 
value of a stream is another stream.  This allows 
us to define \emph{incremental streaming computations for recursive programs}.  As a 
consequence we derive a universal algorithm for incrementalizing arbitrary streaming Datalog programs.

\dbsp is a \textbf{simple} language: 
the basic  \dbsp streaming model is built essentially from two elementary mathematical operators: lifting $\lift$ and delay $\zm$.
The nested streams model adds two additional operators, $\delta_0$ for stream construction and $\int$ for destruction.

\dbsp is also \textbf{expressive}: for example, it is more powerful than stratified Datalog.  
\dbsp can also describe streaming window queries, or queries on nested relations (such as grouping),
and non-monotone recursive queries. 
We discuss briefly the application of \dbsp to richer languages in \secref{sec:extensions}.

This paper omits most proofs; the full proofs are available in an expansive  
companion technical report~\cite{tr}.

This paper makes the following contributions:
\begin{enumerate}[nosep, leftmargin=0pt, itemindent=0.5cm, label=\textbf{(\arabic{*})}]
    \item It defines \dbsp, a small language for streaming computation, which
    nonetheless can express nested non-monotonic recursion;
    \item It provides an algorithm incrementalizing all \dbsp programs;
    \item For fragments of \dbsp corresponding to the relational algebra and stratified-monotonic 
    Datalog, 
    the automatic incrementalization algorithm provides results matching state-of-the-art approaches.
    Moreover, our approach also applies to more powerful languages, such as while-relational and non-monotonic 
    Datalog~\cite{Abiteboul-book95}.
    \item It develops a formal, sound foundation for the manipulation of streaming and incremental computations,
    which allows one to reason formally about program transformations and design new efficient implementations.
    \item \dbsp can express both streaming and incremental computation models in a single framework.  We regard this unification as a significant contribution.
\end{enumerate}

%% file: streams.tex
\section{Stream computations}\label{sec:streams}

In this section we introduce formally the notion of a stream as an
infinite sequence of values, and we define computations on streams.
Stream operators (\secref{sec:notation}) are the basic building block of stream computations.  
We employ (\secref{sec:abelian}) restricted types of stream operators: causal operators (which cannot 
``look into the future''), and strict operators (which cannot even ``look into the present'').
Moreover, all our operators are ``synchronous'': they consume and produce data at the same ``rate''. Causal operators can be chained
into complex acyclic computational circuits; cyclic circuits are restricted to using strict operators 
on back-edges.  Finally, we define (\secref{sec:abelianstreams}) two useful stream operators: integration and differentiation.

All the results in this section have been known for decades, 
but we recapitulate them to clarify our model's formal assumptions.

\subsection{Streams and stream operators}\label{sec:notation}

$\N$ is the set of natural numbers, $\B$ is the set of Booleans, and $\Z$ is the set of integers.

\begin{definition}[stream]
Given a set $A$, a \defined{stream} \emph{of values from $A$}, or an \emph{$A$-stream}, is a function $\N \rightarrow A$. 
We denote by $\stream{A} \defn \{ s \,|\, s : \N \to A \}$ the set of all $A$-streams. 
\end{definition}

When $s\in\stream{A}$ and $t\in\N$ we 
write $s[t]$ for the $t$-th element of the stream $s$ instead of the usual $s(t)$
to distinguish it from other function applications.
We think of the index $t\in\N$ as (discrete) time and of $s[t]\in A$ 
as the value of the the stream $s$ ``at time'' $t$.
\ifstreamexamples
For example, the stream of natural numbers $id \in \stream{\N}$ given by $\id[t] = t$ is the sequence of values
$\sv{0 1 2 3 4}$.
\fi

\begin{definition}[stream operator]
A (typed) \defined{stream operator} with $n$ inputs is a function $T:\stream{A_0}\times\cdots\times\stream{A_{n-1}}\to\stream{B}$. 
\end{definition}

In general we will use ``operator'' for functions on streams, and
``function'' for computations on ``scalar'' values.

We are using an extension of the simply-typed lambda calculus to write \dbsp programs;
we will introduce its elements gradually.  However, we find it more readable to
also use signal-processing-like circuit diagrams to depict \dbsp programs.
In a circuit diagram a rectangle represents an operator application (labeled
with the operator name, e.g., $T$), while an arrow is a stream.  

Stream operator \emph{composition} (function composition) is shown as chained circuits.
The composition of a binary operator $T: \stream{A} \times \stream{B} \to \stream{A}$ with the 
unary operator $S: \stream{A} \to \stream{B}$ into the computation 
$\lambda s. T(T(s,S(s)),S(s)) : \stream{A}\to\stream{A}$ 
is:

\begin{center}
\begin{tikzpicture}[auto,>=latex]
  \node[] (input) {$s$};
  \node[] [right of=input] (dummy) {};
  \node[block, below of=dummy, node distance=.7cm] (S1) {$S$};
  \node[block, right of=S1] (T1) {$T$};
  \node[block, right of=T1] (T2) {$T$};
  \node[block, above of=T2, node distance=.7cm] (S2) {$S$};
  \node[right of=T2] (output) {$o$}; 
  \draw[->] (input) -| (S1);
  \draw[->] (input) -| (T1);
  \draw[->] (S1) -- (T1);
  \draw[->] (T1) -- (T2);
  \draw[->] (input) |- (S2);  \draw[->] (T2) -- (output);
  \draw[->] (S2) -- (T2);
\end{tikzpicture}
\end{center}

(Diagrams obscure the \emph{order} of the inputs of an operator; for non-commutative
operators we have to provide more information.)  

\begin{definition}(lifting)
Given a (scalar) function $f: A \to B$,
we define a stream operator $\lift{f} :\stream{A} \to \stream{B}$ 
by \emph{lifting} the function $f$ pointwise in time: $(\lift{f})(s) \defn f \circ s$.
Equivalently, $(\lift{f})(s)[t] \defn f(s[t])$.
This extends to functions of multiple arguments.
\end{definition}

\ifstreamexamples
For example, $(\lift{(\lambda x.(2x))})(id) = \sv{0 2 4 6 8}$.
\fi

\begin{proposition}[distributivity]\label{prop:distributivity}
Lifting distributes over function composition:
$\lift{(f \circ g)} = (\lift{f}) \circ (\lift{g})$.
\end{proposition}

We say that two \dbsp programs are \defined{equivalent} if they compute the same
input-output function on streams.
We use the symbol $\cong$ to indicate that two circuits are 
equivalent.  For example, Proposition~\ref{prop:distributivity}
states the following circuit equivalence:

\noindent
\begin{tabular}{m{3.5cm}m{.3cm}m{3.5cm}}
\begin{tikzpicture}[auto,>=latex]
  \node[] (input) {$s$};
  \node[block, right of=input] (g) {$\lift{g}$};
  \node[block, right of=g] (f) {$\lift{f}$};
  \node[right of=f] (output) {$o$};
  \draw[->] (input) -- (g);
  \draw[->] (g) -- (f);
  \draw[->] (f) -- (output);
\end{tikzpicture}
&
$\cong$
&
\begin{tikzpicture}[auto,>=latex]
    \node[] (input) {$s$};
    \node[block, right of=input, node distance=1.5cm] (fg) {$\lift{(f \circ g)}$};
    \node[right of=fg, node distance=1.5cm] (output) {$o$};
    \draw[->] (input) -- (fg);
    \draw[->] (fg) -- (output);
\end{tikzpicture}
\end{tabular}

\subsection{Streams over abelian groups}\label{sec:abelian}

For the rest of the technical development we require the set of values $A$
of a stream $\stream{A}$ to form a commutative group $(A, +, 0, -)$. 
Now we introduce the primitive stream operators that \dbsp uses.

\subsubsection{Delays and time-invariance}\label{sec:delay}

\begin{definition}[Delay]
The \defined{delay operator}\footnote{The name $\zm$
comes from the DSP literature, and is related to the z-transform~\cite{rabiner-book75}.}
produces an output stream
by delaying its input by one step: $\zm_A: \stream{A} \to \stream{A}$:

\begin{tabular}{m{5cm}m{3cm}}
$
\zm_A(s)[t] \defn   \begin{cases}
0        & \text{when}~t=0_A \\
s[t - 1] & \text{when}~t\geq1
\end{cases}
$ &
\begin{tikzpicture}[auto,node distance=1cm,>=latex]
    \node[] (input) {$s$};
    \node[block, right of=input] (z) {$\zm$};
    \node[right of=z] (output) {$o$};
    \draw[->] (input) -- (z);
    \draw[->] (z) -- (output);
\end{tikzpicture}
\end{tabular} 
\end{definition}

We often omit the type parameter $A$, and write just $\zm$.
\ifstreamexamples
For example, $\zm(\id) = \sv{0 0 1 2 3}$.
\fi

\begin{definition}[Time invariance]
A stream operator $S: \stream{A} \to \stream{B}$ is \defined{time-invariant} iff 
$S(\zm_A(s)) = \zm_B(S(s))$ for all $s \in \stream{A}$, or, in other words, iff the
two following circuits are equivalent:

\begin{tabular}{m{3cm}m{.5cm}m{3cm}}
\begin{tikzpicture}[auto,>=latex]
  \node[] (input) {$s$};
  \node[block, right of=input] (S) {$S$};
  \node[block, right of=S] (z) {$\zm$};
  \node[right of=z] (output) {$o$};
  \draw[->] (input) -- (S);
  \draw[->] (S) -- (z);
  \draw[->] (z) -- (output);
\end{tikzpicture}
&
$\cong$
&
\begin{tikzpicture}[auto,>=latex]
  \node[] (input) {$s$};
  \node[block, right of=input] (z) {$\zm$};
  \node[block, right of=z] (S) {$S$};
  \node[right of=S] (output) {$o$};
  \draw[->] (input) -- (z);
  \draw[->] (z) -- (S);
  \draw[->] (S) -- (output);
\end{tikzpicture}
\end{tabular}

\noindent
This definition extends 
naturally to operators with multiple inputs.
\end{definition}

The composition of time-invariant operators of any number of inputs
is time invariant. The delay operator $\zm$ is time-invariant. 
\dbsp only uses time-invariant operators.

\begin{definition}
We say that a function between groups $f: A \to B$ has the \emph{zero-preservation
property} if $f(0_A) = 0_B$.  We write $\zpp{f}$.
\end{definition}

A lifted operator $\lift{f}$ is time-invariant iff $\zpp{f}$.

\subsubsection{Causal and strict operators}\label{sec:causal}

\begin{definition}[Causality]
A stream operator $S:\stream{A}\to\stream{B}$
is \defined{causal} when for all $s,s'\in\stream{A}$,
and all times $t$ we have:
$
(\forall i \leq t~s[i]=s'[i]) ~~\Rightarrow~~ S(s)[t]=S(s')[t].
$
\end{definition}

\noindent
In other words, the output value at time $t$ can only depend on 
input values from times $t' \leq t$.
Operators produced by lifting are causal, and $\zm$ is causal.
All \dbsp operators are causal.  The composition
of causal operators of any number of inputs is causal.

\begin{definition}[Strictness]
A stream operator, $F:\stream{A}\to\stream{B}$ 
is \defined{strictly causal} (abbreviated \textbf{strict})
if  $\forall s,s'\in\stream{A}, \forall t \in \N$ we have:
$(\forall i<t . ~s[i]=s'[i]) ~~\Rightarrow~~ F(s)[t]=F(s')[t].$
\end{definition}

So the $t$-th output of $F(s)$ can depend only on ``past'' values
of the input $s$, between $0$ and $t-1$.  
In particular, $F(s)[0] = 0_B$ is the same for all $s \in \stream{A}$.
Strict operators are causal. Lifted operators in general are \emph{not} strict. 
$\zm$ is strict.  

\begin{proposition}
\label{prop-unique-fix}
For a strict $F: \stream{A} \to \stream{A}$ the equation ~$\alpha=F(\alpha)$~ has a unique
solution $\alpha \in \stream{A}$, denoted by $\fix{\alpha}{F(\alpha)}$. 
\end{proposition}

Thus every strict operator from a set to itself has a unique fixed point.
The simple proof relies on strong induction, showing that $\alpha[t]$ 
depends only on the values of $\alpha$ prior to $t$.  

We show that the following circuit, having a strict ``feedback'' edge $F$,
is a well-defined function on streams:

\begin{center}
\begin{tikzpicture}[>=latex]
    \node[] (input) {$s$};
    \node[block, right of=input] (f) {$T$};
    \node[right of=f] (output) {$\alpha$};
    \node[block, below of=f, node distance=.5cm] (z) {$F$};
    \draw[->] (input) -- (f);
    \draw[->] (f) -- node (mid) {} (output);
    \draw[->] (mid.center) |-  (z);
    \draw[->] (z.west) -- ++(-.3,0) |- ([yshift=1mm]f.south west);
\end{tikzpicture}
\end{center}

\begin{lemma} 
\label{lemma-causal-strict}
If $F: \stream{B} \to \stream{B}$ is strict and $T: \stream{A} \times \stream{B} \to \stream{B}$ is causal, then for a fixed $s$ the operator
$\lambda\alpha.T(s,F(\alpha)): \stream{A} \to \stream{B}$ is strict. 
\end{lemma}

\begin{corollary}\label{feedback-semantics}
\label{cor-loop}
If $F: \stream{B} \to \stream{B}$ is strict and $T: \stream{A} \times \stream{B} \to \stream{B}$ is causal,
the operator $Q(s)=\fix{\alpha}{T(s,F(\alpha))}$ is well-defined and causal. 
If, moreover, $F$ and $T$ are time-invariant then so is $Q$.
\end{corollary}

All stream computations in \dbsp are built from the primitive operators 
we have described: lifted operators and delays (we add two more operators in \secref{sec:nested}).
Circuits composed of such operators can be efficiently implemented 
using Dataflow machines~\cite{lee-ieee95}.

Circuits with feedback are used for two purposes: defining an
integration operator (in the next section), and defining recursive computations (\secref{sec:recursion}).  In turn, the
integration operator will be instrumental in defining incremental
computations (\secref{sec:incremental}).

\subsection{Integration and differentiation}\label{sec:abelianstreams}

Remember that we require the elements of a stream to come from an abelian group $A$.  
Streams themselves form an abelian group:

\begin{proposition}
The structure $(\stream{A},+,0,-)$, obtained by lifting the $+$ and unary $-$ operations from $A$
to $\stream{A}$, is an abelian group.
\end{proposition}

\noindent
Stream addition and negation are causal, time-invariant operators.

\begin{definition}
Given abelian groups $A$ and $B$ we call a stream operator 
$S: \stream{A} \rightarrow \stream{B}$ \defined{linear} if it is a group homomorphism, that is,
$S(a+b)=S(a)+S(b)$ (and therefore $S(0)=0$ and $S(-a)=-S(a)$). 
\end{definition}

Lifting a linear function $f: A \to B$ produces
a stream operator $\lift{f}$ that is linear, time-invariant (LTI).
$\zm$ is LTI.   

\begin{definition}(bilinear)
A function of two arguments $f: A \times B \to C$ with $A, B, C$ groups, is \emph{bilinear} 
if it is linear separately in each argument (i.e., it distributes over addition): 
$\forall a, b, c, d . f(a+b, c) = f(a, c) + f(b, c)$, and $f(a, c+d) = f(a, c) + f(c, d).$
\end{definition}

This definition extends to stream operators.
Lifting a bilinear function $f: A \times B \to C$ produces
a bilinear stream operator $\lift{f}$.  An example bilinear operator over $\stream{\Z}$
is lifted multiplication:
$f: \stream{\N} \times \stream{\N} \to \stream{\N}, f(a, b)[t] = a[t]\cdot b[t]$.  

The composition of (bi)linear operators with linear operators 
is (bi)linear (since homomorphisms compose).

The feedback loop produced with a linear operator is linear:

\begin{proposition}
\label{prop-rec-linear}
Let $S$ be a unary causal LTI operator. The
operator $Q(s)=\fix{\alpha}{S(s+\zm(\alpha))}$ is well-defined and LTI:

\begin{center}
\begin{tikzpicture}[>=latex]
    \node[] (input) {$s$};
    \node[block, shape=circle, right of=input, inner sep=0pt, node distance=.6cm] (plus) {$+$};
    \node[block, right of=plus] (Q) {$S$};
    \node[right of=Q, node distance=1.2cm] (output) {$\alpha$};
    \node[block, below of=Q, node distance=.6cm] (z) {$\zm$};
    \draw[->] (input) -- (plus);
    \draw[->] (plus) -- (Q);
    \draw[->] (Q) -- node (mid) {} (output);
    \draw[->] (mid.center) |-  (z);
    \draw[->] (z) -| (plus);
\end{tikzpicture}
\end{center}
\end{proposition}

\begin{definition}[Differentiation]
The \defined{differentiation operator} $\D_{\stream{A}} : \stream{A} \to \stream{A}$ is defined by:
$\D(s) \defn s - \zm(s)$.
\end{definition}
We generally omit the type, and write just $\D$ when the type can be inferred from the context.
The value of $\D(s)$ at time $t$ is the difference
between  the current (time $t$) value of $s$ and the previous (time $t-1$) value of $s$.
\ifstreamexamples
As an example, $\D(\id) = \sv{0 1 1 1 1}$.
\fi

If $s$ is a stream, then $\D(s)$ is the \emph{stream of changes} of $s$. 

\begin{proposition}
\label{prop-diff-properties}
Differentiation $\D$ is causal and LTI.
\end{proposition}

\begin{tabular}{cc}
\begin{tikzpicture}[auto,>=latex,node distance=1cm]
    \node[] (input) {$s$};
    \node[block, shape=circle, right of=input, inner sep=0pt,node distance=2cm] (plus) {$+$};
    \node[right of=plus] (output) {$\D(s)$};
    \draw[draw,->] (input) -- node (i) {} (plus);
    \node[block, below of=i, node distance=.7cm] (z) {$\zm$};
    \node[block, shape=circle, right of=z, inner sep=1pt] (minus) {$-$};
    \draw[->] (plus) -- (output);
    \draw[->] (i) -- (z);
    \draw[->] (z) -- (minus);
    \draw[->] (minus) -- (plus);
\end{tikzpicture} &
\begin{tikzpicture}[auto,>=latex]
    \node[] (input) {$s$};
    \node[block, shape=circle, right of=input, inner sep=0pt] (plus) {$+$};
    \node[right of=plus, node distance=1.5cm] (output) {$\I(s)$};
    \node[block, below of=plus, node distance=.7cm] (z) {$z^{-1}$};
    \draw[draw,->] (input) -- (plus);
    \draw[->] (plus) -- node (o) {} (output);
    \draw[->] (o) |- (z);
    \draw[->] (z) -- (plus);
\end{tikzpicture} \\
Differentiation & Integration
\end{tabular}

The integration operator ``reconstitutes'' a stream from its changes:

\begin{definition}[Integration]
The \defined{integration operator}  $\I_{\stream{A}} : \stream{A} \to \stream{A}$ 
is defined by $\I(s) \defn \lambda s . \fix{\alpha}{(s + \zm(\alpha))}$.
\end{definition}

\noindent
We also generally omit the type, and write just $\I$.
This is the construction from Proposition~\ref{prop-rec-linear} 
using the identity function for $S$.

\begin{proposition}
$\I(s)$ is the discrete (indefinite) integral applied to the stream $s$: 
$\I(s)[t] = \sum_{i \leq t} s[i]$.
\end{proposition}
\ifstreamexamples
As an example, $\I(\id) = \sv{0 1 3 6 10}$.
\fi

\begin{proposition}
\label{prop-integ-properties}
$\I$ is causal and LTI.
\end{proposition}

\begin{theorem}[Inversion]
\label{inverses}
Integration and differentiation are inverses of each other:
$\forall s . \I(\D(s)) = \D(\I(s)) = s$.
\end{theorem}

\noindent
\begin{tabular}{m{2.5cm}m{.3cm}m{1cm}m{.3cm}m{2.5cm}}
\begin{tikzpicture}[auto,>=latex, node distance=.75cm]
    \node[] (input) {$s$};
    \node[block, right of=input] (I) {$\I$};
    \node[block, right of=I] (D) {$\D$};
    \node[right of=D] (output) {$o$};
    \draw[->] (input) -- (I);
    \draw[->] (I) -- (D);
    \draw[->] (D) -- (output);
\end{tikzpicture}
     &  
     $\cong$
     &
     \hspace{-2ex}
\begin{tikzpicture}[auto,>=latex, node distance=.75cm]
    \node[] (input) {$s$};
    \node[right of=input] (output) {$o$};
    \draw[->] (input) -- (output);
\end{tikzpicture}
     &
     $\cong$
     &
\begin{tikzpicture}[auto,>=latex, node distance=.75cm]
    \node[] (input) {$s$};
    \node[block, right of=input] (D) {$\D$};
    \node[block, right of=D] (I) {$\I$};
    \node[right of=I] (output) {$o$};
    \draw[->] (input) -- (D);
    \draw[->] (D) -- (I);
    \draw[->] (I) -- (output);
\end{tikzpicture}
\end{tabular}

\section{Incremental computation}\label{sec:incremental}
 
\begin{definition}
Given a unary stream operator $Q: \stream{A} \to \stream{B}$ we define the 
\defined{incremental version} of $Q$ as $\inc{Q} \defn \D \circ Q \circ \I$.
$\inc{Q}$ has the same ``type'' as $Q$: $\inc{Q}: \stream{A} \to \stream{B}$.
For an operator with multiple inputs we define 
the incremental version by applying $\I$ to each input independently:
e.g., if $T: \stream{A} \times \stream{B} \rightarrow \stream{C}$ then
$\inc{T}(a, b) \defn \D (T(\I(a), \I(b)))$.
\end{definition}

The following diagram illustrates the intuition behind this definition: 
\begin{tikzpicture}[auto,>=latex]
    \node[] (input) {$\Delta s$};
    \node[block, right of=input] (I) {$\I$};
    \node[block, right of=I] (Q) {$Q$};
    \node[block, right of=Q] (D) {$\D$};
    \node[right of=D] (output) {$\Delta o$};
    \draw[->] (input) -- (I);
    \draw[->] (I) -- node (s) {$s$} (Q);
    \draw[->] (Q) -- node (o) {$o$} (D);
    \draw[->] (D) -- (output);
\end{tikzpicture}

If $Q(s) = o$ is a computation, then $\inc{Q}$ performs 
the ``same'' computation as $Q$,
but between streams of changes $\Delta s$ and $\Delta o$.
This is the diagram from the
introduction, substituting $\Delta s$ for the transaction stream $T$, 
and $o$ for the stream of view versions $V$.

Notice that our definition of incremental computation is meaningful only for \emph{streaming}
computations; this is in contrast to classic definitions, e.g.~\cite{gupta-idb95} which
consider only one change.  Generalizing the definition to operate on streams gives us
additional power, especially when operating with recursive queries.

The following proposition is one of our central results.

\begin{proposition}(Properties of the incremental version):
For computations of appropriate types, the following hold:
\label{prop-inc-properties}
\begin{description}[nosep, leftmargin=\parindent]
\item[inversion:] $Q\mapsto\inc{Q}$ is bijective; its inverse is $Q\mapsto \I\circ Q\circ\D$.
\item[invariance:] $\inc{+} = +, \inc{(\zm)} = \zm, \inc{-} = -, \inc{\I}=\I, \inc{\D}=\D$
\item[push/pull:] \label{prop-part-commutation}
    $Q \circ \I = \I \circ \inc{Q}$; $\D\circ Q = \inc{Q}\circ\D$
\item[chain:] $\inc{(Q_1\circ Q_2)} = \inc{Q_1}\circ\inc{Q_2}$ (This generalizes to operators with multiple inputs.)
\item[add:] $\inc{(Q_1 + Q_2)} = \inc{Q_1} + \inc{Q_2}$
\item[cycle:] $\inc{(\lambda s. \fix{\alpha}{T(s,\zm(\alpha)}))} = \lambda s. \fix{\alpha}{\inc{T}(s,\zm(\alpha)})$
\end{description}
\end{proposition}

The proof of these properties relies on elementary algebraic manipulations.
Despite their simplicity, they are very useful.
For example, the \defined{chain rule}  states that the following two circuits are equivalent:

\noindent
\begin{tabular}{m{4cm}m{.2cm}m{2.5cm}}
\begin{tikzpicture}[auto,>=latex,node distance=.8cm]
  \node[] (input) {$i$};
  \node[block, right of=input] (I) {$\I$};
  \node[block, right of=I] (Q1) {$Q_1$};
  \node[block, right of=Q1] (Q2) {$Q_2$};
  \node[block, right of=Q2] (D) {$\D$};
  \node[right of=D] (output)  {$o$};
  \draw[->] (input) -- (I);
  \draw[->] (I) -- (Q1);
  \draw[->] (Q1) -- (Q2);
  \draw[->] (Q2) -- (D);
  \draw[->] (D) -- (output);
\end{tikzpicture} &
$\cong$ &
\begin{tikzpicture}[>=latex]
  \node[] (input) {$i$};
  \node[block, right of=input] (Q1) {$\inc{Q_1}$};
  \node[block, right of=Q1] (Q2) {$\inc{Q_2}$};
  \node[right of=Q2] (output)  {$o$};
  \draw[->] (input) -- (Q1);
  \draw[->] (Q1) -- (Q2);
  \draw[->] (Q2) -- (output);
\end{tikzpicture}
\end{tabular}

\noindent In other words, \textbf{to incrementalize a composite query you can incrementalize
each sub-query independently}.  This gives us a simple deterministic recipe  
for computing the incremental version of an arbitrarily complex query.

We illustrate by giving the proof of the 
chain rule, which is trivial, and is based on 
function composition associativity:
$$
\begin{aligned}
\inc{(Q_1 \circ Q_2)} &= \D \circ Q_1 \circ Q_2 \circ \I \\
 &= \D \circ Q_1 \circ (\I \circ \D) \circ Q_2 \circ \I \\
 &= (\D \circ Q_1 \circ \I) \circ (\D \circ Q_2 \circ \I) \\
 &= \inc{(Q_1)} \circ \inc{(Q_2)}.
\end{aligned}
$$

The \defined{cycle rule} states that the following circuits are equivalent:

\noindent
\begin{tabular}{m{4.2cm}m{.2cm}m{3cm}}
\begin{tikzpicture}[>=latex]
    \node[] (input) {$s$};
    \node[block, right of=input] (I) {$\I$};
    \node[block, right of=I] (f) {$T$};
    \node[block, right of=f] (D) {$\D$};
    \node[right of=D] (output) {$o$};
    \node[block, below of=f, node distance=.7cm] (z) {$\zm$};
    \draw[->] (input) -- (I);
    \draw[->] (I) -- (f);
    \draw[->] (f) -- node (mid) {} (D);
    \draw[->] (mid.center) |-  (z);
    \draw[->] (z.west) -- ++(-.3,0) |- ([yshift=1mm]f.south west);
    \draw[->] (D) -- (output);
\end{tikzpicture} & $\cong$ &
\begin{tikzpicture}[>=latex]
    \node[] (input) {$s$};
    \node[block, right of=input] (f) {$\inc{T}$};
    \node[right of=f, node distance=1.5cm] (output) {$o$};
    \node[block, below of=f, node distance=.7cm] (z) {$\zm$};
    \draw[->] (input) -- (f);
    \draw[->] (f) -- node (mid) {} (output);
    \draw[->] (mid.center) |-  (z);
    \draw[->] (z.west) -- ++(-.3,0) |- ([yshift=1mm]f.south west);
\end{tikzpicture}
\end{tabular}

The incremental version of a feedback loop around a query
is just the feedback loop with the incremental query.  The significance
of this result will be apparent when we implement recursive queries.

To execute incremental queries efficiently, we want to compute directly 
on streams of changes without integrating them. The invariance property above shows
that stream operators $+$, $-$, and $\zm$ are identical to their incremental versions, 
thus $\I$ and $\D$ can be omitted for them: $\inc{Q} = \I\circ Q\circ\D = Q$.
The following theorems generalize this to linear and bi-linear operators:

\begin{theorem}[Linear]\label{linear}
For an LTI operator $Q$ we have $\inc{Q}=Q$.
\end{theorem}

\begin{theorem}[Bilinear]\label{bilinear}
For a bilinear time-invariant operator $\times$ we have
$\inc{(a \times b)} ~=~ a \times b ~+~ \zm(\I(a)) \times b ~+~ a \times \zm(\I(b))$.
\end{theorem}

By rewriting this statement using $\Delta a$ for the stream of changes to $a$ we
get the familiar formula for incremental equi-joins:
$\Delta(a\times b) =\Delta a \times \Delta b + a\times(\Delta b) + (\Delta a)\times b$.

This should not be surprising because equi-joins are bilinear,
as we discuss in the next section.  

%% file: relational.tex
\section{Incremental View Maintenance}\label{sec:relational}

Results in \secref{sec:streams} and~\secref{sec:incremental}
apply to streams of arbitrary group values.  In this
section we turn our attention to using these results in the context of 
relational view maintenance.  As explained in the introduction, we want to
efficiently compute the incremental version of any relational query $Q$
that updates a database view.

However, we face a technical problem: the $\I$ and $\D$ operators were
defined on abelian groups, and relational databases in general are 
not abelian groups, since they operate on sets.  Fortunately, 
there is a well-known tool in the database literature
which converts set operations into group operations by using \zrs
(also called z-relations~\cite{green-pods07}) instead of sets.

We start by defining the \zrs group, and then we review how 
relational queries are converted into \dbsp circuits  over \zrs.  
What makes this translation efficiently incrementalizable is the fact that
many basic relational queries can be expressed using LTI \zr operators.

\subsection{\zrs as an abelian group}

Given a set $A$, we define \defined{\zrs}\footnote{Also called $\Z$-relations
elsewhere~\cite{green-tcs11}, because often A is a Cartesian product in practice;
however, we only need the set structure for most of our results.}
over $A$ as functions with \emph{finite support} from $A$ to $\Z$.  
These are functions $f: A \rightarrow \Z$ where 
$f(x) \not= 0$ for at most a finite number of values $x \in A$.
We also write $\Z[A]$ for the type of \zrs with elements from $A$.
Values in $\Z[A]$ can be thought of as key-value maps with 
keys in $A$ and values in $\Z$, justifying the array indexing notation.
We write $f[a]$ instead of $f(a)$.
Since $\Z$ is an abelian ring, $\Z[A]$ is also an abelian ring (and thus a group).  This group
$(\Z[A], +_{\Z[A]}, 0_{\Z[A]}, -_{\Z{A}})$ has addition and subtraction defined pointwise: 
$(f +_{\Z[A]} g)(x) = f(x) + g(x) . \forall x \in A.$  
The $0$ element of $\Z[A]$ is the function $0_{\Z[A]}$ defined by $0_{\Z[A]}(x) = 0 . 
\forall x \in A$.

A particular \zr $m \in \Z[A]$ can be denoted by enumerating the
inputs that map to non-zero values and their corresponding values: 
$m = \{ x_1 \mapsto w_1, \dots, x_n \mapsto w_n \}$.  
We call $w_i \in Z$ the \defined{multiplicity} (or weight)
of $x_i \in A$.  Multiplicities can be negative.  
We write that $x \in m$ for $x \in A$, iff $m[x] \not= 0$.

\ifzsetexamples
For example, let's consider a concrete \zr $R \in \Z[\texttt{string}]$,
defined by $R = \{ \texttt{joe} \mapsto 1, \texttt{anne} \mapsto -1 \}$.  
$R$ has two elements in its domain,
\texttt{joe} with a multiplicity of 1 (so $R[\texttt{joe}] = 1$), 
and \texttt{anne} with a multiplicity of $-1$.
We say \texttt{joe} $\in R$ and \texttt{anne} $\in R$.
\fi

\zrs generalize sets and bags.  A set with elements from $A$
can be represented as a \zr by associating a weight of 1 with each set element.
When translating queries on sets to \dbsp programs we 
convert the data values back and forth between sets and \zrs.

\begin{definition}
We say that a \zr represents a \defined{set} if the multiplicity of every
element is one.  We define a function to check this property 
$\isset : \Z[A] \rightarrow \B$\index{isset}
given by:
$$\isset(m) \defn \left\{
\begin{array}{ll}
  \mbox{true} & \mbox{ if } m[x] = 1, \forall x \in m \\
  \mbox{false} & \mbox{ otherwise}
\end{array}
\right.
$$
\end{definition}

\ifzsetexamples
For our example $\isset(R) = \mbox{false}$, since $R[\texttt{anne}] = -1$.  
\fi

\begin{definition}
We say that a \zr is \defined{positive} (or a \defined{bag}) if the multiplicity of every element is
positive.  We define a function to check this property
$\ispositive : \Z[A] \rightarrow \B$\index{ispositive}.
given by
$$\ispositive(m) \defn \left\{
\begin{array}{ll}
  \mbox{true} & \mbox{ if } m[x] \geq 0, \forall x \in A \\
  \mbox{false} & \mbox{ otherwise}
\end{array}
\right.$$
$\forall m \in \Z[A] . \isset(m) \Rightarrow \ispositive(m)$.
\end{definition}

\ifzsetexamples
We have $\ispositive(R) = \mbox{false}$, since $R[\texttt{anne}] = -1$.
\fi

We write $m \geq 0$ when $m$ is
positive.  For positive $m, n$ we write $m \geq n$ for $m, n
\in \Z[A]$ iff $m - n \geq 0$.  $\geq$ is a partial order.

We call a function $f : \Z[A] \rightarrow \Z[B]$ \defined{positive} if it maps
positive values to positive values:
$\forall x \in \Z[A], x \geq 0_{\Z[A]} \Rightarrow f(x) \geq 0_{\Z[B]}$.  
We apply this notation to functions as well: $\ispositive(f)$.

\begin{definition}[distinct]
The function $\distinct: \Z[A] \rightarrow \Z[A]$\index{distinct}
projects a \zr into an underlying set (but \emph{the result is
  still a \zr}):
$$\distinct(m)[x] \defn \left\{
\begin{array}{ll}
  1 & \mbox{ if } m[x] > 0 \\
  0 & \mbox{ otherwise}
\end{array}
\right.
$$
\end{definition}

$\distinct$ ``removes'' elements with negative multiplicities.
\ifzsetexamples
$\distinct(R) = \{ \texttt{joe} \mapsto 1 \}$.
\fi

While very simple, this definition of $\distinct$ has been carefully 
chosen to enable us to define precisely all relational (set) operators
from \zrs operators.

Circuits derived from relational queries only compute on positive \zrs; 
negative values will only be used to represent \emph{changes} to \zrs.
Negative weights ``remove'' elements from a set.

All the results from \secref{sec:streams} extend to streams over \zrs. 

\begin{definition}(mononotonicity)
A stream $s \in \stream{\Z[A]}$ is \defined{positive} if every value of the stream is positive:
$s[t] \geq 0 . \forall t \in \N$.
A stream $s \in \stream{\Z[A]}$ is \defined{monotone} if $s[t] \geq s[t-1], \forall t \in \N$.
\end{definition}

If $s \in \stream{\Z[A]}$ is positive, then $\I(s)$ is monotone.
If $s \in \stream{\Z[A]}$ is monotone, $\D(s)$ is positive.

\paragraph{Generalizing box-and-arrow diagrams}

From now on we will use circuits to compute both on scalars and streams.
We use the same graphical representation for functions on streams or scalars: 
boxes with input and output arrows.  For scalar functions the ``values'' 
of the arrows are scalars instead of streams; otherwise
the interpretation of boxes as function application is unchanged.

\subsection{Implementing relational operators}\label{sec:relational-operators}

The fact that relational algebra can be implemented by computations
on \zrs has been shown before, e.g.~\cite{green-pods07}.  The translation
of all the core relational operators is shown in Table~\ref{tab:relational}.  
The translation is essentially given 
by induction on the query structure.

\newlength{\commentsize}
\setlength{\commentsize}{5cm}
\begin{table*}[h]
\small
\begin{tabular}{|m{1.2cm}m{4.2cm}m{5cm}m{\commentsize}|} \hline
Operation & SQL example & \dbsp circuit & Details \\ \hline
Composition &
 \begin{lstlisting}[language=SQL]
SELECT DISTINCT ... FROM 
(SELECT ... FROM ...)
\end{lstlisting}
 & 
 \begin{tikzpicture}[auto,>=latex]
  \node[] (I) {\code{I}};
  \node[block, right of=I] (CI) {$C_I$};
  \draw[->] (I) -- (CI);
  \node[block, right of=CI] (CO) {$C_O$};
  \node[right of=CO] (O) {\code{O}};
  \draw[->] (CI) -- (CO);
  \draw[->] (CO) -- (O); 
\end{tikzpicture}
 &
 \parbox[b][][t]{\commentsize}{
  $C_I$ circuit for inner query, \\
  $C_O$ circuit for outer query.}  
\\ \hline
Union & 
\begin{lstlisting}[language=SQL]
(SELECT * FROM I1) 
UNION 
(SELECT * FROM I2)
\end{lstlisting}
&
\begin{tikzpicture}[auto,>=latex]
  \node[] (input1) {\code{I1}};
  \node[below of=input1, node distance=.4cm] (midway) {};
  \node[below of=midway, node distance=.4cm] (input2) {\code{I2}};
  \node[block, shape=circle, right of=midway, inner sep=0in] (plus) {$+$};
  \node[block, right of=plus, node distance=1.5cm] (distinct) {$\distinct$};
  \node[right of=distinct, node distance=1.5cm] (output) {\code{O}};
  \draw[->] (input1) -| (plus);
  \draw[->] (input2) -| (plus);
  \draw[->] (plus) -- (distinct);
  \draw[->] (distinct) -- (output);
\end{tikzpicture}
&
\\ \hline
Projection &
\begin{lstlisting}[language=SQL]
SELECT DISTINCT I.c 
FROM I
\end{lstlisting}
&
\begin{tikzpicture}[auto,>=latex]
  \node[] (input) {\code{I}};
  \node[block, right of=input] (pi) {$\pi$};
  \node[block, right of=pi, node distance=1.5cm] (distinct) {$\distinct$};
  \node[right of=distinct, node distance=1.5cm] (output) {\code{O}};
  \draw[->] (input) -- (pi);
  \draw[->] (pi) -- (distinct);
  \draw[->] (distinct) -- (output);
\end{tikzpicture}
&
\parbox[b][][t]{\commentsize}{
$\pi(i)[y] \defn
\sum_{x \in i, x|_c = y} i[x]$ \\
$x|_c$ is projection on column $c$ of the tuple $x$ \\
$\pi$ is linear; $\ispositive(\pi), \zpp{\pi}$.
}
\\ \hline
Filtering &
\begin{lstlisting}[language=SQL]
SELECT * FROM I 
WHERE p(I.c)
\end{lstlisting}
&
\begin{tikzpicture}[auto,>=latex]
  \node[] (input) {\code{I}};
  \node[block, right of=input] (map) {$\sigma_P$};
  \node[block, right of=map] (distinct) {$\distinct$};
  \node[right of=distinct] (output) {\code{O}};
  \draw[->] (input) -- (map);
  \draw[->] (map) -- (distinct);
  \draw[->] (distinct) -- (output);
\end{tikzpicture}
&
\parbox[b][][t]{\commentsize}{
$\sigma_P(m)[x] \defn \left\{
\begin{array}{ll}
  m[x] \cdot x & \mbox{ if } P(x) \\
  0 & \mbox{ otherwise } \\
\end{array}
\right.$ \\
$P: A \rightarrow \B$ is a predicate. \\
$\sigma_P$ is linear; $\ispositive(\sigma_P), \zpp{\sigma_P}$.
}
 \\ \hline
Selection &
\begin{lstlisting}[language=SQL]
SELECT DISTINCT f(I.c, ...) 
FROM I
\end{lstlisting}
&
\begin{tikzpicture}[auto,>=latex]
  \node[] (input) {\code{I}};
  \node[block, right of=input, node distance=1.5cm] (map) {$\mbox{map}(f)$};
  \node[block, right of=map, node distance=1.5cm] (distinct) {$\distinct$};
  \node[right of=distinct, node distance=1.5cm] (output) {\code{O}};
  \draw[->] (input) -- (map);
  \draw[->] (map) -- (distinct);
  \draw[->] (distinct) -- (output);
\end{tikzpicture}
& 
\parbox[b][][t]{\commentsize}{
For a function $f$ \\
$\map(f)$ is linear, \\
$\ispositive(\map(f)), \zpp{\map(f)}$
}.
\\ \hline
\parbox[b][][t]{1cm}{
Cartesian \\
product} &
\begin{lstlisting}[language=SQL]
SELECT I1.*, I2.* 
FROM I1, I2
\end{lstlisting}
& 
\begin{tikzpicture}[auto,>=latex]
  \node[] (i1) {\code{I1}};
  \node[below of=i1, node distance=.4cm] (midway) {};
  \node[below of=midway, node distance=.4cm] (i2) {\code{I2}};
  \node[block, right of=midway] (prod) {$\times$};
  \node[right of=prod] (output) {\code{O}};
  \draw[->] (i1) -| (prod);
  \draw[->] (i2) -| (prod);
  \draw[->] (prod) -- (output);
\end{tikzpicture}
& 
\parbox[b][][t]{\commentsize}{
$(a \times b)((x,y)) \defn a[x] \times b[y]$. \\
$\times$ is bilinear, $\ispositive(\times), \zpp{\times}$.
}
\\ \hline
Equi-join &
\begin{lstlisting}[language=SQL]
SELECT I1.*, I2.* 
FROM I1 JOIN I2
ON I1.c1 = I2.c2
\end{lstlisting}
&
\begin{tikzpicture}[auto,>=latex]
  \node[] (i1) {\code{I1}};
  \node[below of=i1, node distance=.4cm] (midway) {};
  \node[below of=midway, node distance=.4cm] (i2) {\code{I2}};
  \node[block, right of=midway] (prod) {$\bowtie$};
  \node[right of=prod] (output) {\code{O}};
  \draw[->] (i1) -| (prod);
  \draw[->] (i2) -| (prod);
  \draw[->] (prod) -- (output);
\end{tikzpicture}
&
\parbox[b][][t]{\commentsize}{
$(a \bowtie b)((x,y)) \defn a[x] \times b[y] \\
\mbox{ if } x|_{c1} = y|_{c2}$. \\
$\bowtie$ is bilinear, $\ispositive(\bowtie), \zpp{\bowtie}$.
}
\\ \hline
Intersection &
\begin{lstlisting}[language=SQL]
(SELECT * FROM I1)
INTERSECT 
(SELECT * FROM I2)
\end{lstlisting}
&
\begin{tikzpicture}[auto,>=latex]
  \node[] (i1) {\code{I1}};
  \node[below of=i1, node distance=.4cm] (midway) {};
  \node[below of=midway, node distance=.4cm] (i2) {\code{I2}};
  \node[block, right of=midway] (prod) {$\bowtie$};
  \node[right of=prod] (output) {\code{O}};
  \draw[->] (i1) -| (prod);
  \draw[->] (i2) -| (prod);
  \draw[->] (prod) -- (output);
\end{tikzpicture}
&
Special case of equi-join when both relations have the same schema.
 \\ \hline
Difference &
\begin{lstlisting}[language=SQL]
SELECT * FROM I1 
EXCEPT 
SELECT * FROM I2
\end{lstlisting}
&
\begin{tikzpicture}[auto,>=latex, node distance=.7cm]
  \node[] (i1) {\code{I1}};
  \node[below of=i1, node distance=.4cm] (midway) {};
  \node[below of=midway, node distance=.4cm] (i2) {\code{I2}};
  \node[block, shape=circle, inner sep=0in, right of=i2] (m) {$-$};
  \node[block, right of=midway, shape=circle, inner sep=0in, node distance=1.3cm] (plus) {$+$};
  \node[block, right of=plus, node distance=1.5cm] (distinct) {$\distinct$};
  \node[right of=distinct, node distance=1.5cm] (output) {\code{O}};
  \draw[->] (i1) -| (plus);
  \draw[->] (i2) -- (m);
  \draw[->] (m) -| (plus);
  \draw[->] (plus) -- (distinct);
  \draw[->] (distinct) -- (output);
\end{tikzpicture}
&
\\ \hline
\end{tabular}
\caption{Implementation of SQL relational set operators in \dbsp.  
Each query assumes that inputs \code{I}, \code{I1}, \code{I2}, are sets and it 
produces output sets.\label{tab:relational}}
\end{table*}

The translation is fairly straightforward, but many operators require
the application of a $\distinct$ to produce sets.  The correctness of
this implementation is predicated on the global circuit inputs being
sets as well.  For example, $a \cup b = \distinct(a + b)$, $a \setminus b = 
\distinct(a - b)$, $(a \times b)((x,y)) = a[x] \times b[y]$.

\subsubsection{Correctness of the \dbsp implementations}\label{sec:correctness}

A relational query $Q$ that transforms
a set $V$ into a set $U$ will be implemented by a \dbsp computation $Q'$ on 
\zrs.  The correctness of the implementation requires that the following
diagram commutes:

\begin{center}
\begin{tikzpicture}
  \node[] (V) {$V$};
  \node[below of=V] (VZ) {$VZ$};
  \node[right of=V, node distance=2cm] (U) {$U$};
  \node[below of=U] (UZ) {$UZ$};
  \draw[->] (V) -- node (f) [above] {$Q$} (U);
  \draw[->] (V) --  node (s) [left] {tozset}(VZ);
  \draw[->] (VZ) -- node (f) [above] {$Q'$} (UZ);
  \draw[->] (UZ) -- node (d) [right] {toset} (U);
\end{tikzpicture}
\end{center}

The $\mbox{toset}$ and $\mbox{tozset}$ functions convert sets to \zrs and 
vice-versa:

$\mbox{toset}: \Z[A] \to 2^A$ is defined by $\mbox{toset}(m) \defn \cup_{x \in \distinct(m)} \{ x \}$.

$\mbox{tozset}: 2^A \to \Z[A]$ is defined by $\mbox{tozset}(s) \defn \sum_{x \in s} 1 \cdot x$.

All standard algebraic properties
of the relational operators can be used to optimize circuits
(they can even be applied to queries before building the circuits).

Notice that the use of the $\distinct$ operator allows DBSP to model
the \emph{full relational algebra}, including difference (and not just the positive
fragment).  Most of the operators that appear in the circuits in
Table~\ref{tab:relational} are linear, and thus have very efficient
incremental versions.  A notable exception is $\distinct$.  While we
show below that $\distinct$ can be computed efficiently incrementally,
it does have an important cost in terms of memory, so we try to 
minimize its use.  For this we can use a pair of optimizations:

\begin{proposition}\label{prop-distinct-delay}
Let Q be one of the following \zrs operators: filtering $\sigma$,
join $\bowtie$, or Cartesian product $\times$.
Then we have $\forall i \in \Z[I], \ispositive(i) \Rightarrow Q(\distinct(i)) = \distinct(Q(i))$.
\end{proposition}

This rule allows us to delay the application of $\distinct$.

\begin{proposition}\label{prop-distinct-once}
Let Q be one of the following \zrs operators: filtering $\sigma$,
projection $\pi$, selection (map($f$)), addition $+$, join $\bowtie$, or
Cartesian product $\times$.
Then we have $\forall i \in \Z[I], \ispositive(i) \Rightarrow \distinct(Q(\distinct(i))) = \distinct(Q(i))$.
\end{proposition}

This is Proposition 6.13 in~\cite{green-tcs11}.

These properties allow us to ``consolidate'' distinct operators by performing
one $\distinct$ at the end of a chain of computations.

Finally, the next proposition shows that the incremental of $\distinct$
can be computed with work proportional to the size of the input change.

\begin{proposition}\label{prop-inc_distinct}
The following circuit implements $\inc{(\lift{\distinct})}$:
\begin{tabular}{m{3.5cm}m{.5cm}m{6cm}}
\begin{tikzpicture}[auto,node distance=1.5cm,>=latex]
    \node[] (input) {$d$};
    \node[block, right of=input] (d) {$\inc{(\lift{\distinct})}$};
    \node[right of=d] (output) {$o$};
    \draw[->] (input) -- (d);
    \draw[->] (d) -- (output);
\end{tikzpicture} &
$\cong$ &
\begin{tikzpicture}[>=latex]
    \node[] (input) {$d$};
    \node[block, right of=input] (I) {$\I$};
    \node[block, right of=I] (z) {$\zm$};
    \node[block, below of=z, node distance=.8cm] (H) {$\lift{H}$};
    \node[right of=H] (output) {$o$};
    \draw[->] (input) -- node (mid) {} (I);
    \draw[->] (I) -- (z);
    \draw[->] (mid.center) |- (H);
    \draw[->] (z) -- node (i) [right] {$i$} (H);
    \draw[->] (H) -- (output);
\end{tikzpicture}
\end{tabular}

\noindent where $H: \Z[A] \times \Z[A] \to \Z[A]$ is defined as:
$$
H(i, d)[x] \defn 
\begin{cases}
-1 & \mbox{if } i[x] > 0 \mbox{ and } (i + d)[x] \leq 0 \\
1  & \mbox{if } i[x] \leq 0 \mbox{ and } (i + d)[x] > 0 \\
0  & \mbox{otherwise} \\
\end{cases}
$$
\end{proposition}

The function $H$ detects whether the multiplicity of an element in the 
input set $i$ when adding change $d$ is changing from negative to
positive or vice-versa.  Notice that only multiplicities of the elements
that appear in the change $d$ can change from input to output, so the
work needed to compute both $H$ is bounded by the size of $d$ and not $i$.

\subsection{Incremental view maintenance}

Let us consider a relational query $Q$ 
defining a view.  To create a circuit that maintains incrementally the view defined by $Q$
we apply the following mechanical steps; this algorithm is deterministic
and its running time is proportional to the complexity 
of the query (number of operators in the query):

\begin{algorithm}[incremental view maintenance]\label{algorithm-inc}\quad
\begin{enumerate}[nosep, leftmargin=\parindent]
    \item Translate $Q$ into a circuit using the rules in Table~\ref{tab:relational}.
    \item Apply $\distinct$ consolidation until convergence.
    \item Lift the whole circuit, by applying Proposition~\ref{prop:distributivity},
    converting it to a circuit operating on streams.
    \item Incrementalize the whole circuit ``surrounding'' it with $\I$ and $\D$.
    \item Apply the chain rule and other properties of the $\inc{\cdot}$ operator
    from Proposition~\ref{prop-inc-properties} recursively on the query structure
    to optimize the incremental implementation.  
\end{enumerate}
\end{algorithm}

Step (3) yields a circuit that consumes a stream of complete database snapshots and outputs a 
stream of complete view snapshots. Step (4) yields a circuit that consumes a stream of changes
to the database and outputs a stream of view changes; however, the internal operation of the 
circuit is non-incremental, as it computes on the complete state of the database reconstructed
by the integration operator.  Step (5) incrementalizes the internals of the circuit by rewriting 
it to compute on changes, avoiding integration when possible (see \secref{sec:incremental}).  

\subsection{Example}\label{sec:relational-example}

In this section we apply the incremental view maintenance algorithm to a concrete
query.  Let us consider the following query:

\begin{lstlisting}[language=SQL]
CREATE VIEW v AS
SELECT DISTINCT t1.x, t2.y FROM (
     SELECT t1.x, t1.id 
     FROM t 
     WHERE t.a > 2 
) t1  
JOIN (
     SELECT t2.id, t2.y
     FROM r 
     WHERE r.s > 5 
) t2 ON t1.id = t2.id
\end{lstlisting}

Step 1: First we create a \dbsp circuit to represent this query using the
translation rules from Table~\ref{tab:relational}:

\noindent
\begin{tikzpicture}[node distance=1.2cm,>=latex]
    \node[] (t1) {\code{t1}};
    \node[block, right of=t1, node distance=.9cm] (s1) {$\sigma_{a > 2}$};
    \node[block, right of=s1] (d1) {$\distinct$};
    \node[block, right of=d1] (p1) {$\pi_{x, d}$};
    \node[block, right of=p1] (d11) {$\distinct$};
    \node[below of=t1, node distance=1cm] (t2) {\code{t2}};
    \node[block, right of=t2, node distance=.9cm] (s2) {$\sigma_{s > 5}$};
    \node[block, right of=s2] (d2) {$\distinct$};
    \node[block, right of=d2] (p2) {$\pi_{y, id}$};
    \node[block, right of=p2] (d21) {$\distinct$};
    \node[below of=d11, node distance=.5cm] (mid) {};
    \node[block, right of=mid, node distance=.8cm] (j) {$\bowtie_{id = id}$};
    \node[block, right of=j] (p) {$\pi_{x, y}$};
    \node[block, right of=p] (d) {$\distinct$};
    \node[right of=d, node distance=.9cm] (V) {\code{V}};
    \draw[->] (t1) -- (s1);
    \draw[->] (s1) -- (d1);
    \draw[->] (d1) -- (p1);
    \draw[->] (p1) -- (d11);
    \draw[->] (t2) -- (s2);
    \draw[->] (s2) -- (d2);
    \draw[->] (d2) -- (p2);
    \draw[->] (p2) -- (d21);
    \draw[->] (d11) -| (j);
    \draw[->] (d21) -| (j);
    \draw[->] (j) -- (p);
    \draw[->] (p) -- (d);
    \draw[->] (d) -- (V);
\end{tikzpicture}

Step 2: we apply the $\distinct$ optimization rules; first the rule from~\ref{prop-distinct-once}
gives us the following equivalent circuit:

\noindent
\begin{tikzpicture}[node distance=1.2cm,>=latex]
    \node[] (t1) {\code{t1}};
    \node[block, right of=t1, node distance=.9cm] (s1) {$\sigma_{a > 2}$};
    \node[block, right of=s1] (p1) {$\pi_{x, d}$};
    \node[block, right of=p1] (d11) {$\distinct$};
    \node[below of=t1, node distance=1cm] (t2) {\code{t2}};
    \node[block, right of=t2, node distance=.9cm] (s2) {$\sigma_{s > 5}$};
    \node[block, right of=s2] (p2) {$\pi_{y, id}$};
    \node[block, right of=p2] (d21) {$\distinct$};
    \node[below of=d11, node distance=.5cm] (mid) {};
    \node[block, right of=mid, node distance=.8cm] (j) {$\bowtie_{id = id}$};
    \node[block, right of=j] (p) {$\pi_{x, y}$};
    \node[block, right of=p] (d) {$\distinct$};
    \node[right of=d, node distance=.9cm] (V) {\code{V}};
    \draw[->] (t1) -- (s1);
    \draw[->] (s1) -- (p1);
    \draw[->] (p1) -- (d11);
    \draw[->] (t2) -- (s2);
    \draw[->] (s2) -- (p2);
    \draw[->] (p2) -- (d21);
    \draw[->] (d11) -| (j);
    \draw[->] (d21) -| (j);
    \draw[->] (j) -- (p);
    \draw[->] (p) -- (d);
    \draw[->] (d) -- (V);
\end{tikzpicture}

Applying the rule from~\ref{prop-distinct-delay} we get:

\noindent
\begin{tikzpicture}[node distance=1.2cm,>=latex]
    \node[] (t1) {\code{t1}};
    \node[block, right of=t1, node distance=.9cm] (s1) {$\sigma_{a > 2}$};
    \node[block, right of=s1] (p1) {$\pi_{x, d}$};
    \node[below of=t1, node distance=1cm] (t2) {\code{t2}};
    \node[block, right of=t2, node distance=.9cm] (s2) {$\sigma_{s > 5}$};
    \node[block, right of=s2] (p2) {$\pi_{y, id}$};
    \node[below of=p1, node distance=.5cm] (mid) {};
    \node[block, right of=mid, node distance=.8cm] (j) {$\bowtie_{id = id}$};
    \node[block, right of=j] (d0) {$\distinct$};
    \node[block, right of=d0] (p) {$\pi_{x, y}$};
    \node[block, right of=p] (d) {$\distinct$};
    \node[right of=d, node distance=.9cm] (V) {\code{V}};
    \draw[->] (t1) -- (s1);
    \draw[->] (s1) -- (p1);
    \draw[->] (t2) -- (s2);
    \draw[->] (s2) -- (p2);
    \draw[->] (p1) -| (j);
    \draw[->] (p2) -| (j);
    \draw[->] (j) -- (d0);
    \draw[->] (d0) -- (p);
    \draw[->] (p) -- (d);
    \draw[->] (d) -- (V);
\end{tikzpicture}

And applying again~\ref{prop-distinct-once} we get:

\noindent
\begin{tikzpicture}[node distance=1.2cm,>=latex]
    \node[] (t1) {\code{t1}};
    \node[block, right of=t1, node distance=.9cm] (s1) {$\sigma_{a > 2}$};
    \node[block, right of=s1] (p1) {$\pi_{x, d}$};
    \node[below of=t1, node distance=1cm] (t2) {\code{t2}};
    \node[block, right of=t2, node distance=.9cm] (s2) {$\sigma_{s > 5}$};
    \node[block, right of=s2] (p2) {$\pi_{y, id}$};
    \node[below of=p1, node distance=.5cm] (mid) {};
    \node[block, right of=mid, node distance=.8cm] (j) {$\bowtie_{id = id}$};
    \node[block, right of=j] (p) {$\pi_{x, y}$};
    \node[block, right of=p] (d) {$\distinct$};
    \node[right of=d, node distance=.9cm] (V) {\code{V}};
    \draw[->] (t1) -- (s1);
    \draw[->] (s1) -- (p1);
    \draw[->] (t2) -- (s2);
    \draw[->] (s2) -- (p2);
    \draw[->] (p1) -| (j);
    \draw[->] (p2) -| (j);
    \draw[->] (j) -- (p);
    \draw[->] (p) -- (d);
    \draw[->] (d) -- (V);
\end{tikzpicture}

Step 3: we lift the circuit using distributivity of composition over lifting; we
obtain a circuit that computes over streams, i.e., for each new input pair of relations
\code{t1} and \code{t2} it will produce an output view \code{V}:

\noindent
\begin{tikzpicture}[node distance=1.3cm,>=latex]
    \node[] (t1) {\code{t1}};
    \node[block, right of=t1, node distance=.9cm] (s1) {$\lift{\sigma_{a > 2}}$};
    \node[block, right of=s1] (p1) {$\lift{\pi_{x, d}}$};
    \node[below of=t1, node distance=1.2cm] (t2) {\code{t2}};
    \node[block, right of=t2, node distance=.9cm] (s2) {$\lift{\sigma_{s > 5}}$};
    \node[block, right of=s2] (p2) {$\lift{\pi_{y, id}}$};
    \node[below of=p1, node distance=.6cm] (mid) {};
    \node[block, right of=mid, node distance=.8cm] (j) {$\lift{\bowtie_{id = id}}$};
    \node[block, right of=j] (p) {$\lift{\pi_{x, y}}$};
    \node[block, right of=p] (d) {$\lift{\distinct}$};
    \node[right of=d] (V) {\code{V}};
    \draw[->] (t1) -- (s1);
    \draw[->] (s1) -- (p1);
    \draw[->] (t2) -- (s2);
    \draw[->] (s2) -- (p2);
    \draw[->] (p1) -| (j);
    \draw[->] (p2) -| (j);
    \draw[->] (j) -- (p);
    \draw[->] (p) -- (d);
    \draw[->] (d) -- (V);
\end{tikzpicture}

Step 4: incrementalize circuit, obtaining a circuit that computes over changes;
this circuit receives changes to relations \code{t1} and \code{t2} and for each
such change it produces the corresponding change in the output view \code{V}:

\noindent
\begin{tikzpicture}[node distance=1.3cm,>=latex]
    \node[] (t1) {$\Delta$\code{t1}};
    \node[block, right of=t1, node distance=.8cm] (I1) {$\I$};
    \node[block, right of=I1, node distance=.9cm] (s1) {$\lift{\sigma_{a > 2}}$};
    \node[block, right of=s1] (p1) {$\lift{\pi_{x, d}}$};
    \node[below of=t1, node distance=1.2cm] (t2) {$\Delta$\code{t2}};
    \node[block, right of=t2, node distance=.8cm] (I2) {$\I$};
    \node[block, right of=I2, node distance=.9cm] (s2) {$\lift{\sigma_{s > 5}}$};
    \node[block, right of=s2] (p2) {$\lift{\pi_{y, id}}$};
    \node[below of=p1, node distance=.6cm] (mid) {};
    \node[block, right of=mid, node distance=.6cm] (j) {$\lift{\bowtie_{id = id}}$};
    \node[block, right of=j] (p) {$\lift{\pi_{x, y}}$};
    \node[block, right of=p] (d) {$\lift{\distinct}$};
    \node[block, right of=d, node distance=1.1cm] (D) {$\D$};
    \node[right of=D, node distance=.7cm] (V) {$\Delta$\code{V}};
    \draw[->] (t1) -- (I1);
    \draw[->] (I1) -- (s1);
    \draw[->] (s1) -- (p1);
    \draw[->] (t2) -- (I2);
    \draw[->] (I2) -- (s2);
    \draw[->] (s2) -- (p2);
    \draw[->] (p1) -| (j);
    \draw[->] (p2) -| (j);
    \draw[->] (j) -- (p);
    \draw[->] (p) -- (d);
    \draw[->] (d) -- (D);
    \draw[->] (D) -- (V);
\end{tikzpicture}

Step 5: apply the chain rule to rewrite the circuit as a composition of incremental operators;

\noindent
\begin{tikzpicture}[node distance=1.6cm,>=latex]
    \node[] (t1) {$\Delta$\code{t1}};
    \node[block, right of=t1, node distance=1.2cm] (s1) {$\inc{(\lift{\sigma_{a > 2}})}$};
    \node[block, right of=s1] (p1) {$\inc{(\lift{\pi_{x, d}})}$};
    \node[below of=t1, node distance=1.2cm] (t2) {$\Delta$\code{t2}};
    \node[block, right of=t2, node distance=1.2cm] (s2) {$\inc{(\lift{\sigma_{s > 5}})}$};
    \node[block, right of=s2] (p2) {$\inc{(\lift{\pi_{y, id}})}$};
    \node[below of=p1, node distance=.6cm] (mid) {};
    \node[block, right of=mid, node distance=.8cm] (j) {$\inc{(\lift{\bowtie_{id = id}})}$};
    \node[block, right of=j] (p) {$\inc{(\lift{\pi_{x, y}})}$};
    \node[block, right of=p] (d) {$\inc{(\lift{\distinct})}$};
    \node[right of=d, node distance=1.2cm] (V) {$\Delta$\code{V}};.8
    \draw[->] (t1) -- (s1);
    \draw[->] (s1) -- (p1);
    \draw[->] (t2) -- (s2);
    \draw[->] (s2) -- (p2);
    \draw[->] (p1) -| (j);
    \draw[->] (p2) -| (j);
    \draw[->] (j) -- (p);
    \draw[->] (p) -- (d);
    \draw[->] (d) -- (V);
\end{tikzpicture}

Use the linearity of $\sigma$ and $\pi$ to simplify this circuit:

\noindent
\begin{tikzpicture}[node distance=1.6cm,>=latex]
    \node[] (t1) {$\Delta$\code{t1}};
    \node[block, right of=t1, node distance=1cm] (s1) {$\lift{\sigma_{a > 2}}$};
    \node[block, right of=s1] (p1) {$\lift{\pi_{x, d}}$};
    \node[below of=t1, node distance=1.2cm] (t2) {$\Delta$\code{t2}};
    \node[block, right of=t2, node distance=1cm] (s2) {$\lift{\sigma_{s > 5}}$};
    \node[block, right of=s2] (p2) {$\lift{\pi_{y, id}}$};
    \node[below of=p1, node distance=.6cm] (mid) {};
    \node[block, right of=mid, node distance=.8cm] (j) {$\inc{(\lift{\bowtie_{id = id}})}$};
    \node[block, right of=j] (p) {$\lift{\pi_{x, y}}$};
    \node[block, right of=p] (d) {$\inc{(\lift{\distinct})}$};
    \node[right of=d, node distance=1.3cm] (V) {$\Delta$\code{V}};.8
    \draw[->] (t1) -- (s1);
    \draw[->] (s1) -- (p1);
    \draw[->] (t2) -- (s2);
    \draw[->] (s2) -- (p2);
    \draw[->] (p1) -| (j);
    \draw[->] (p2) -| (j);
    \draw[->] (j) -- (p);
    \draw[->] (p) -- (d);
    \draw[->] (d) -- (V);
\end{tikzpicture}

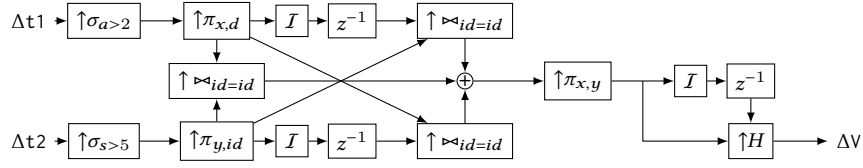
\begin{figure*}[h]
\begin{tikzpicture}[node distance=1.5cm,>=latex]
    \node[] (t1) {$\Delta$\code{t1}};
    \node[block, right of=t1, node distance=1cm] (s1) {$\lift{\sigma_{a > 2}}$};
    \node[block, right of=s1] (p1) {$\lift{\pi_{x, d}}$};
    \node[below of=t1, node distance=1.6cm] (t2) {$\Delta$\code{t2}};
    \node[block, right of=t2, node distance=1cm] (s2) {$\lift{\sigma_{s > 5}}$};
    \node[block, right of=s2] (p2) {$\lift{\pi_{y, id}}$};
    
      \node[block, right of=p1, node distance=1cm] (jI1) {$\I$};
      \node[block, below of=p1, node distance=.8cm] (ab) {$\lift\bowtie_{id=id}$};
      \node[block, right of=p2, node distance=1cm] (jI2) {$\I$};
      \draw[->] (p1) -- (jI1);
      \draw[->] (p2) -- (jI2);
      \node[block, right of=jI1, node distance=.8cm] (ZI1) {$\zm$};
      \node[block, right of=jI2, node distance=.8cm] (ZI2) {$\zm$};
      \draw[->] (jI1) -- (ZI1);
      \draw[->] (jI2) -- (ZI2);
      \node[block, right of=ZI1] (DI1) {$\lift\bowtie_{id=id}$};
      \node[block, right of=ZI2] (DI2) {$\lift\bowtie_{id=id}$};
      \draw[->] (ZI1) -- (DI1);
      \draw[->] (ZI2) -- (DI2);
      \node[block, circle, below of=DI1, inner sep=0cm, node distance=.8cm] (sum) {$+$};
      \draw[->] (ab) -- (sum);
      \draw[->] (DI1) -- (sum);
      \draw[->] (DI2) -- (sum);
      \draw[->] (p1) -- (ab);
      \draw[->] (p2) -- (ab);
      \draw[->] (p1) -- (DI2);
      \draw[->] (p2) -- (DI1);
    
    \node[block, right of=sum] (p) {$\lift{\pi_{x, y}}$};
    \draw[->] (sum) -- (p);
    \node[block, right of=p] (Id) {$\I$};
    \node[block, right of=Id, node distance=.8cm] (zd) {$\zm$};
    \node[block, below of=zd, node distance=.8cm] (H) {$\lift{H}$};
    \node[right of=H, node distance=1.3cm] (V) {$\Delta$\code{V}};.8
    \draw[->] (t1) -- (s1);
    \draw[->] (s1) -- (p1);
    \draw[->] (t2) -- (s2);
    \draw[->] (s2) -- (p2);
    \draw[->] (p) -- node (tapp) {} (Id);
    \draw[->] (Id) -- (zd);
    \draw[->] (zd) -- (H);
    \draw[->] (tapp.center) |- (H);
    \draw[->] (H) -- (V);
\end{tikzpicture}
\caption{Final version of the incremental query circuit from 
\refsec{sec:relational-example}\label{fig:relational-example}.}
\end{figure*}

Finally, replace the incremental join using the formula for bilinear operators 
(Theorem~\ref{bilinear}),
and the incremental $\distinct$ (Proposition~\ref{prop-inc_distinct}),
obtaining the circuit from Figure~\ref{fig:relational-example}.

Notice that the resulting circuit contains three integration operations: two from
the join, and one from the $\distinct$.  It also contains three join operators.
However, the work performed by each operator
for each new input is proportional to the size of change, as we argue in the following section.

\subsection{Complexity}

Incremental circuits are efficient.  The work performed (and the memory used) by a circuit
is the sum of the work performed (and memory used) by its operators.
We argue that each operator in the incremental version of a circuit is efficient.

For incrementalized circuits the input stream of each operator contains \emph{changes} in
its input relations.  Denote $C[t] \defn \norm{s[t]}$ the \defined{size}
of the value of stream $s$ of changes at time $t$, and $R[t] \defn
\norm{\I(s)[t]}$ the size of the relation produced by integrating all changes in
$s$.  An unoptimized incremental operator $\inc{Q} = \D \circ Q \circ \I$
evaluates query $Q$ on the integration of its input streams; hence its time
complexity  is the same as that of the non-incremental operator, a function of
$R[t]$.  In addition, because of the $\I$ and $\D$ operators, it uses $O(R[t])$ memory.

The optimizations described in \secref{sec:incremental} reduce the
reduce the time complexity of an operator to be a function of $C[t]$.  Assuming $C[t] \ll
R[t]$, this translates to major performance improvements in practice.  For
example, Theorem~\ref{linear}, allows evaluating $\inc{T}$, where $T$ is a
linear operator, in time $O(C[t])$.  Interestingly, while the $\I$
operator uses $O(R[t])$ memory, it can be evaluated in $O(C[t])$ time, because 
all values that appear in the output at time $t$ must be present in
current input change for time $t$.  Similarly, while the $\distinct$ operator is not
linear, $\inc{(\lift{\distinct})}$ can also be evaluated in $O(C[t])$ according to
Proposition~\ref{prop-inc_distinct}.  Bilinear operators, including join, can be
evaluated in time proportional to the product of the sizes of their input
changes $O(C[t]^2)$ (Theorem~\ref{bilinear}).  

The space complexity of linear operators is 0 (zero), since they store no
data persistently.  The space complexity of $\inc{(\lift{\distinct})}$ and 
join is $O(R[t])$.

%% file: recursion.tex
\section{Recursive queries}\label{sec:recursion}

Recursive queries are very useful in a many applications.
For example, many graph algorithms (such as graph reachability
or transitive closure) are naturally expressed using recursive queries.

We introduce two new stream operators that are instrumental in
expressing recursive query evaluation.  These operators allow us
to build circuits implementing looping constructs, which 
are used to iterate computations until a fixed-point is reached.

\begin{definition}\label{def:zae}
We say that a stream $s \in \stream{A}$ is \defined{zero almost-everywhere} if it has a finite 
number of non-zero values, i.e., there exists a time $t_0 \in \N$
s.t. $\forall t \geq t_0 . s[t] = 0$.
\noindent Denote the set of streams that are zero almost everywhere
by $\streamf{A}$.
\end{definition}

\paragraph{Stream introduction}

The delta function (named from the Dirac delta function) $\delta_0 : A \rightarrow \stream{A}$
produces a stream from a scalar value:
$$\delta_0(v)[t] \defn \left\{
\begin{array}{ll}
  v & \mbox{if } t = 0 \\
  0_A & \mbox{ otherwise}
\end{array}
\right.
$$
\ifstreamexamples
For example, $\delta_0(5)$ is the stream $\sv{5 0 0 0 0}$.
\fi

\paragraph{Stream elimination}

We define the function $\int : \streamf{A} \rightarrow
A$, over streams that are zero almost everywhere, as 
$\int(s) \defn \sum_{t \geq 0} s[t]$.  
$\int$ is closely related to $\I$; if $\I$ is the
indefinite integral, $\int$ is the definite integral on the
interval $0 - \infty$.

For many implementation strategies (including  
relational and Datalog queries given below) the $\int$
operator can be approximated finitely and precisely by integrating until the
first 0 value encountered, since it can be proven that its input is always
the derivative of a monotone stream.

$\delta_0$ is the left inverse of $\int$, i.e.: $\int \circ \; \delta_0 = \id_A$.  
\begin{proposition}
$\delta_0$ and $\int$ are LTI.
\end{proposition}

\paragraph{Nested time domains}

So far we used a tacit assumption that ``time'' is common for all
streams in a program.  For example, when we add two streams, 
we assume that they use the same ``clock'' for the time dimension.
However, the $\delta_0$ operator creates a stream with a ``new'', independent time
dimension.  We require \emph{well-formed circuits}
to ``insulate'' such
nested time domains by nesting them between a $\delta_0$ and an $\int$ operator:

\begin{center}
\begin{tikzpicture}[auto,node distance=1cm,>=latex]
    \node[] (input) {$i$};
    \node[block, right of=input] (delta) {$\delta_0$};
    \node[block, right of=delta] (f) {$Q$};
    \draw[->] (input) -- (delta);
    \draw[->] (delta) -- (f);

    \node[block, right of=f] (S) {$\int$};
    \node[right of=S] (output) {$o$};
    \draw[->] (f) -- (S);
    \draw[->] (S) -- (output);
\end{tikzpicture}
\end{center}

\begin{proposition}
If $Q$ is time-invariant, the circuit above has the zero-preservation
property: $\zpp{\int \circ\; Q \circ \delta_o}$.
\end{proposition}

\subsection{Implementing Recursive Datalog}\label{sec:datalog}

We illustrate the implementation of recursive queries in \dbsp for
stratified Datalog.
Datalog is strictly more expressive than
relational algebra since it can express recursive programs, e.g.:
{\small
\begin{lstlisting}[language=ddlog]
O(v) :- I(v).  // base case
O(v) :- I(z), O(x), v = ... .  // rec case
\end{lstlisting}
}
In general, a recursive Datalog program defines a set of
mutually recursive relations $O_1,..,O_n$ as an equation
$(O_1,..,O_n)=R(I_1,..,I_m, O_1,..,O_n)$, where $I_1,..,I_m$ are
input relations and $R$ is a relational (non-recursive) query.

The following algorithm generates a circuit that computes a
solution to this equation.  We describe the algorithm informally and 
for the special case of a single input $I$ and single output $O$; the general case
can be found in the companion technical report~\cite{tr}, and is only 
slightly more involved.

\noindent 
\begin{enumerate}[nosep, leftmargin=\parindent]
\item Implement the non-recursive relational query $R$ as described in
    \secref{sec:relational} and Table~\ref{tab:relational}; this produces
    an acyclic circuit whose inputs and outputs are a \zr (i.e., not a stream):
    \begin{center}
    \begin{tikzpicture}[auto,>=latex]
      \node[] (I) {\code{I}};
      \node[below of=I, node distance=.5cm] (O) {\code{O}};
      \node[block, right of=I] (R) {$R$};
      \node[right of=R] (o) {\code{O}};
      \draw[->] (I) -- (R);
      \draw[->] (O) -| (R);
      \draw[->] (R) -- (o);
    \end{tikzpicture} 
    \end{center}
\item Lift this circuit to operate on streams and connect the output to the input in a feedback cycle as follows:

\noindent
\begin{tikzpicture}[auto,>=latex, node distance=.8cm]
  \node[] (Iinput) {\code{I}};
  \node[block, right of=Iinput] (ID) {$\delta_0$};
  \node[block, right of=ID] (II) {$\I$};
  \node[block, right of=II] (f) {$\lift{R}$};
  \node[block, right of=f, node distance=1.5cm] (distinct) {$\lift{\distinct}$};
  \node[block, right of=distinct, node distance=1.5cm] (D) {$\D$};
  \node[block, right of=D] (S) {$\int$};
  \node[right of=S] (output)  {\code{O}};
  \draw[->] (Iinput) -- (ID);
  \draw[->] (ID) -- (II);
  \draw[->] (II) -- (f);
  \draw[->] (f) -- (distinct);
  \draw[->] (distinct) -- node (o) {$o$} (D);
  \draw[->] (D) -- (S);
  \draw[->] (S) -- (output);
  \node[block, below of=distinct, node distance=.7cm] (z) {$\zm$};
  \draw[->] (o) |- (z);
  \draw[->] (z) -| (f);
\end{tikzpicture}

We construct $\lift{R}$ by lifting each operator of the circuit individually 
according to Proposition~\ref{prop:distributivity}.
\end{enumerate}

The inner loop of the circuit computes the fixed point of $R$.  The differentiation
operator $\D$ yields the set of new Datalog facts (changes) computed by each iteration of the loop.
When the set of new facts becomes empty the iterations have completed.
$\int$ computes the value of the fixed point by aggregating these changes.

\begin{theorem}[Recursion correctness]\label{theorem:recursion}
If $\isset(\code{I})$, the output of the circuit above is
the relation $\code{O}$ as defined by the Datalog semantics 
as a function of the input relation \code{I}.
\end{theorem}
\label{proof-recursion}
\begin{proof}
Let us compute the contents of the $o$ stream, produced at the output
of the $\distinct$ operator.  We will show that this stream is composed
of increasing approximations of the value of \code{O}.

We define the following one-argument function: $S(x) = \lambda x . R(\code{I}, x)$.
Notice that the left input of the $\lift{R}$ block is a constant stream
with the value \code{I}.  Due to the stratified nature of the language,
we must have $\ispositive(S)$, so $\forall x . S(x) \geq x$.
Also $\lift{S}$ is time-invariant, so $S(0) = 0$.
From \secref{sec:relational}, the definition of set union we know that
$x \cup y = \distinct(x + y)$.
We get the following system of equations:
$$
\begin{aligned}
o[0] =& S(0) \\
o[t] =& S(o[t-1]) \\
\end{aligned}
$$ 
So, by induction on $t$ we have $o[t] = S^t(0)$, where by 
$S^t$ we mean $\underbrace{S \circ S \circ \ldots \circ S}_{t}$.
$S$ is monotone; thus, if there is a time $k$ such that $S^k(0) = S^{k+1}(0)$, we have 
$\forall j \in \N . S^{k+j}(0) = S^k(0)$.  Applying a derivative to this stream
will then produce a stream that is zero almost everywhere, and integrating
this derivative will return the last distinct value in the stream $o$.

This is essentially the definition of the semantics of a recursive Datalog relation:
$\code{O} = \fix{x}{R(\code{I}, x)}$.
\end{proof}

Note that the use of unbounded data domains (like integers with arithmetic) does 
not guarantee convergence for all programs.

In fact, this circuit implements the standard \defined{na\"{\i}ve evaluation}
algorithm (e.g., see Algorithm~1 in \cite{greco-sldm15}).
Notice that the inner part of the circuit is the incremental
form of another circuit, since it is sandwiched between $\I$ and $\D$ operators.
Using the cycle rule of Proposition~\ref{prop-inc-properties} we can rewrite this circuit as:
\begin{equation}
\begin{aligned}
\label{eq:seminaive}
\begin{tikzpicture}[auto,>=latex]
  \node[] (Iinput) {\code{I}};
  \node[block, right of=Iinput] (Idelta) {$\delta_0$};
  \node[block, right of=Idelta] (f) {$\inc{(\lift{R})}$};
  \node[block, right of=f, node distance=1.6cm] (distinct) {$\inc{\lift{(\distinct})}$};
  \node[block, right of=distinct, node distance=1.5cm] (S) {$\int$};
  \node[right of=S] (output)  {\code{O}};
  \node[block, below of=distinct, node distance=.7cm] (z) {$\zm$};
  \draw[->] (Iinput) -- (Idelta);
  \draw[->] (f) -- (distinct);
  \draw[->] (distinct) -- node (o) {} (S);
  \draw[->] (S) -- (output);
  \draw[->] (o) |- (z);
  \draw[->] (z) -| (f);
  \draw[->] (Idelta) -- (f);
\end{tikzpicture}
\end{aligned}
\end{equation}

This last circuit effectively implements the \defined{semi-na\"{\i}ve evaluation}
algorithm (Algorithm~2 from~\cite{greco-sldm15}).  The correctness of semi-na\"{\i}ve
evaluation is an immediate consequence of the cycle rule.

%% file: nested.tex
\section{Incremental recursive programs}\label{sec:nested}

In \secref{sec:streams}--\ref{sec:relational} 
we showed how to incrementalize a relational query by
compiling it into a circuit, lifting the circuit to compute on streams, and
applying the $\inc{\cdot}$ operator to the lifted circuit.  In \secref{sec:datalog} we showed
how to compile a recursive query into a circuit that employs incremental
computation internally to compute the fixed point.
Here we combine these results to construct a circuit that evaluates a \emph{recursive
query incrementally}.  The circuit receives a stream of updates to input
relations, and for every update recomputes the fixed point.  To do this
incrementally, it preserves the stream of changes to recursive relations
produced by the iterative fixed point computation, and adjusts this stream to
account for the modified inputs.  Thus, every element of the input stream yields
a stream of adjustments to the fixed point computation, using
\emph{nested streams}.

Nested streams, or streams of streams, $\stream{\stream{A}} = \N \rightarrow (\N
\rightarrow A)$, are well defined, since streams form an abelian group.
Equivalently, a nested stream is a value in $\N \times \N \to A$, i.e., a matrix
with an infinite number of rows, indexed by two-dimensional time $(t_0, t_1)$. 
where each row is a stream.  In
\secref{sec-nested-examples} we show a few example nested stream
computations.

Lifting a stream operator $S: \stream{A} \to \stream{B}$ yields an operator over
nested streams $\lift{S}: \stream{\stream{A}} \to \stream{\stream{B}}$, such
that $(\lift{S})(s) = S \circ s$, or, pointwise: $(\lift{S}(s))[t_0][t_1] =
S(s[t_0])[t_1], \forall t_0, t_1 \in \N$.  In particular, a scalar function $f:
A \rightarrow B$ can be lifted twice to produce an operator between streams of
streams: $\lift{\lift{f}}: \stream{\stream{A}} \rightarrow \stream{\stream{B}}$.

We define a partial order over timestamps: $(i_0, i_1)
\leq (t_0, t_1)$ iff $i_0 \leq t_0$ and $i_1 \leq t_1$.  We extend the
definition of strictness for operators over nested streams: a stream operator
$F: \stream{\stream{A}} \to \stream{\stream{B}}$ is strict if for any $s, s' \in
\stream{\stream{A}}$ and all times $t, i \in \N \times \N$ we have $\forall i <
t, s[i] = s'[i]$ implies $F(s)[t] = F(s')[t]$.
Proposition~\ref{prop-unique-fix} holds for this notion of strictness, i.e., the fixed point operator $\fix{\alpha}{F(\alpha)}$ is well defined for a strict operator $F$.

\begin{proposition}\label{prop-liftz}
The operator $\lift{\zm}: \stream{\stream{A}} \to \stream{\stream{A}}$ is strict.
\end{proposition}

The operator $\zm$ on nested streams delays ``rows'' of the matrix, 
while $\lift{\zm}$ delays ``columns''.  (See examples in 
\secref{sec-nested-examples}).

The $\I$ operator on $\stream{\stream{A}}$ operates on rows
of the matrix, treating each row as a single value.
Lifting a stream operator computing on $\stream{A}$, 
such as $\I: \stream{A} \to \stream{A}$, also produces an operator on nested streams, but
this time computing on the columns of the matrix
$\lift{\I}: \stream{\stream{A}} \to \stream{\stream{A}}.$  

\begin{proposition}[Lifting cycles]
\label{prop-lift-cycle}
For a binary, causal $T$ we have:
$\lift{(\lambda s. \fix{\alpha}{T(s,\zm(\alpha)}))} = \lambda s. \fix{\alpha}{(\lift{T})(s,(\lift{\zm})(\alpha))}$
\noindent i.e., lifting a circuit containing a ``cycle'' can be accomplished by
lifting all operators independently, including the $\zm$ back-edge.
\end{proposition}

This means that lifting a \dbsp stream function can be expressed within \dbsp
itself.  For example, we have:

\begin{tabular}{m{2cm}m{.5cm}m{4cm}}
\begin{tikzpicture}[>=latex]
  \node[] (input) {$i$};
  \node[block, right of=input] (I) {$\lift{\I}$};
  \node[right of=I] (output)  {$o$};
  \draw[->] (input) -- (I);
  \draw[->] (I) -- (output);
\end{tikzpicture}
& $\cong$ &
\begin{tikzpicture}[>=latex]
  \node[] (input) {$i$};
  \node[block, circle, right of=input, inner sep=0cm] (p) {$+$};
  \node[right of=p, node distance=1.5cm] (output)  {$o$};
  \node[block, below of=p, node distance=.7cm] (z) {$\lift{\zm}$};
  \draw[->] (input) -- (p);
  \draw[->] (p) -- node (mid) {} (output);
  \draw[->] (z) -- (p);
  \draw[->] (mid.center) |- (z);
\end{tikzpicture}
\end{tabular}

This proposition gives the ability to lift
entire circuits, including circuits computing on streams and having feedback edges,
which are well-defined, due to Proposition~\ref{prop-liftz}.  
With this machinery we can now apply Algorithm~\ref{algorithm-inc} to arbitrary
circuits, even circuits built for recursively-defined relations.  
Consider the ``semi-naive'' circuit~(\ref{eq:seminaive}),
and denote $\distinct \circ R$ with $T$:

\begin{center}
\begin{tikzpicture}[>=latex]
  \node[] (Iinput) {\code{I}};
  \node[block, right of=Iinput] (Idelta) {$\delta_0$};
  \node[block, right of=Idelta] (f) {$\inc{(\lift{T})}$};
  \node[block, right of=f, node distance=1.5cm] (S) {$\int$};
  \node[right of=S] (output)  {\code{O}};
  \draw[->] (f) -- node (o) {} (S);
  \node[block, below of=o, node distance=.7cm] (z) {$\zm$};
  \draw[->] (Iinput) -- (Idelta);
  \draw[->] (S) -- (output);
  \draw[->] (o.center) -- (z);
  \draw[->] (z) -| (f);
  \draw[->] (Idelta) -- (f);
\end{tikzpicture}
\vspace{-2mm}
\end{center}

\noindent Lift the entire circuit using Proposition~\ref{prop-lift-cycle} and incrementalize it:

\begin{tikzpicture}[>=latex]
  \node[] (Iinput) {\code{I}};
  \node[block, right of=Iinput] (I) {$\I$};
  \node[block, right of=I] (Idelta) {$\lift{\delta_0}$};
  \node[block, right of=Idelta, node distance=1.5cm] (f) {$\lift{\inc{(\lift{T})}}$};
  \node[block, right of=f, node distance=1.5cm] (S) {$\lift{\int}$};
  \node[block, right of=S] (D) {$\D$};
  \node[right of=D] (output)  {\code{O}};
  \draw[->] (f) -- node (o) {} (S);
  \node[block, below of=o, node distance=.7cm] (z) {$\lift{\zm}$};
  \draw[->] (Iinput) -- (I);
  \draw[->] (I) -- (Idelta);
  \draw[->] (S) -- (D);
  \draw[->] (D) -- (output);
  \draw[->] (o.center) -- (z);
  \draw[->] (z) -| (f);
  \draw[->] (Idelta) -- (f);
\end{tikzpicture}

\noindent Now apply the chain rule to this circuit:
\begin{equation}
\vspace{-2.1ex}
\begin{aligned}
\label{eq:increcursive}
\begin{tikzpicture}[>=latex]
  \node[] (Iinput) {\code{I}};
  \node[block, right of=Iinput] (Idelta) {$\lift{\delta_0}$};
  \node[block, right of=Idelta, node distance=2cm] (f) {$\inc{(\lift{\inc{(\lift{T})}})}$};
  \node[block, right of=f, node distance=2cm] (S) {$\lift{\int}$};
  \node[right of=S] (output)  {\code{O}};
  \draw[->] (f) -- node (o) {} (S);
  \node[block, below of=o, node distance=.7cm] (z) {$\lift{\zm}$};
  \draw[->] (Iinput) -- (Idelta);
  \draw[->] (S) -- (output);
  \draw[->] (o.center) -- (z);
  \draw[->] (z) -| (f);
  \draw[->] (Idelta) -- (f);
\end{tikzpicture}
\end{aligned}
\end{equation}
This is the incremental version of an arbitrary recursive query.

\subsection{Example}\label{sec:recursive-example}

In this section we derive the incremental version of a circuit containing
recursion, by applying Algorithm~\ref{algorithm-inc}.  We start with a very simple
program, expressed in Datalog, which computes the transitive closure of a directed
graph:

\begin{lstlisting}[language=ddlog]
// Edge relation with head and tail
input relation E(h: Node, t: Node)
// Reach relation with source s and sink t
output relation R(s: Node, t: Node)
R(x, x) :- E(x, _).
R(x, x) :- E(_, x).
R(x, y) :- E(x, y).
R(x, y) :- E(x, z), R(z, y).
\end{lstlisting}

We haven't explained how Datalog is translated to circuits, but most Datalog
operators are relational in nature.  Assuming one could write 
recursive queries in SQL where a view is defined in terms of itself, the
above program would be implemented by the following (illegal) SQL query:

\begin{lstlisting}[language=SQL]
CREATE VIEW R AS 
(SELECT E.h, E.h FROM E)
UNION
(SELECT E.t, E.t FROM E)
UNION
(SELECT * FROM E)
UNION
(SELECT E.h, R.t 
 FROM E JOIN R 
 ON E.t = R.s)
\end{lstlisting}

We apply the algorithm from \refsec{sec:datalog} to create first the non-recursive circuit,
by assuming that \code{R} is already computed as a view \code{R1}, and using \code{R1}
in the definition of \code{R} instead of itself:

\begin{lstlisting}[language=SQL]
CREATE VIEW Reach AS 
(SELECT E.h, E.h FROM E)
UNION
(SELECT E.t, E.t FROM E)
UNION
(SELECT * FROM E)
UNION
(SELECT E.h, R1.t 
 FROM E JOIN R ON E.t = R1.s)
\end{lstlisting}

Now we implement this query as a \dbsp circuit with two inputs \code{E} and \code{R1}:

\noindent
\begin{tikzpicture}[>=latex, node distance=1.2cm]
  \node[] (E) {\code{E}};
  \node[above of=E, node distance=.6cm] (R1) {\code{R1}};
  \node[block, right of=R1] (j) {$\bowtie_{t=s}$};
  \node[block, right of=j] (pj) {$\pi_{h, t}$};
  \node[block, below of=j] (p1) {$\pi_{h}$};
  \node[block, right of=p1] (s1) {$\sigma_{h, h}$};
  \node[block, below of=p1, node distance=.6cm] (p2)  {$\pi_{t}$};
  \node[block, right of=p2] (s2) {$\sigma_{t, t}$};
  \node[below of=pj, node distance=.6cm] (mid) {};
  \node[block, circle, right of=mid, inner sep=0cm, node distance=1.5cm] (plus) {$+$};
  \node[block, right of=plus] (d) {$\distinct$};
  \node[right of=d] (R) {\code{R}};
  \draw[->] (R1) -- (j);
  \draw[->] (E) -- (j);
  \draw[->] (j) -- (pj);
  \draw[->] (E) -- (p1);
  \draw[->] (p1) -- (s1);
  \draw[->] (E) -- (p2);
  \draw[->] (p2) -- (s2);
  \draw[->] (E) -- (plus);
  \draw[->] (pj) -- (plus);
  \draw[->] (s1) -- (plus);
  \draw[->] (s2) -- (plus);
  \draw[->] (plus) -- (d);
  \draw[->] (d) -- (R);
\end{tikzpicture}

Now lift the circuit by lifting each operator pointwise, and connect it in a feedback
loop by connecting input \code{R1} from the output \code{R} through a \zm operator and bracket
everything with $\delta_0 -- \int$:

\noindent
\begin{tikzpicture}[>=latex]
  \node[] (Einput) {\code{E}};
  \node[block, right of=Einput, node distance=.8cm] (ID) {$\delta_0$};
  \node[block, right of=ID, node distance=.8cm] (E) {$\I$};
  
  \node[right of=E] (empty) {};
  \node[block, above of=empty, node distance=.6cm] (j) {$\lift{\bowtie_{t=s}}$};
  \node[block, right of=j, node distance=1.2cm] (pj) {$\lift{\pi_{h, t}}$};
  \node[block, below of=j] (p1) {$\lift{\pi_{h}}$};
  \node[block, right of=p1] (s1) {$\lift{\sigma_{h, h}}$};
  \node[block, below of=p1, node distance=.6cm] (p2)  {$\lift{\pi_{t}}$};
  \node[block, right of=p2] (s2) {$\lift{\sigma_{t, t}}$};
  \node[below of=pj, node distance=.6cm] (mid) {};
  \node[block, circle, right of=mid, inner sep=0cm, node distance=1.2cm] (plus) {$+$};
  \node[block, right of=plus] (d) {$\lift{\distinct}$};
  \draw[->] (E) -- (j);
  \draw[->] (j) -- (pj);
  \draw[->] (E) -- (p1);
  \draw[->] (p1) -- (s1);
  \draw[->] (E) -- (p2);
  \draw[->] (p2) -- (s2);
  \draw[->] (E) -- (plus);
  \draw[->] (pj) -- (plus);
  \draw[->] (s1) -- (plus);
  \draw[->] (s2) -- (plus);
  \draw[->] (plus) -- (d);
  
  \node[block, right of=d, node distance=1.1cm] (D) {$\D$};
  \node[block, right of=D, node distance=.8cm] (S) {$\int$};
  \node[right of=S, node distance=.8cm] (output)  {\code{R}};
  \draw[->] (Einput) -- (ID);
  \draw[->] (ID) -- (E);
  \draw[->] (d) -- node (o) {} (D);
  \draw[->] (D) -- (S);
  \draw[->] (S) -- (output);
  \node[block, above of=j, node distance=.8cm] (z) {$\zm$};
  \draw[->] (o.center) |- (z);
  \draw[->] (z) -- (j);
\end{tikzpicture}

The above circuit is a complete implementation of the non-streaming
recursive query; given an input relation \code{E} it will produce
its transitive closure \code{R} at the output.  

Now we use the semina\"ive property~\ref{eq:seminaive} to rewrite the circuit:

(To save space in the figures we will omit the indices from $\pi$ and $\sigma$
in the subsequent figures, for example by writing just $\pi$ instead of $\pi_h$.)

\noindent
\begin{tikzpicture}[>=latex, node distance=1.4cm]
  \node[] (Einput) {\code{E}};
  \node[block, right of=Einput, node distance=.8cm] (E) {$\delta_0$};
  
  \node[right of=E] (empty) {};
  \node[block, above of=empty, node distance=.6cm] (j) {$\inc{(\lift{\bowtie})}$};
  \node[block, right of=j] (pj) {$\inc{(\lift{\pi})}$};
  \node[block, below of=j, node distance=1cm] (p1) {$\inc{(\lift{\pi})}$};
  \node[block, right of=p1] (s1) {$\inc{(\lift{\sigma)}}$};
  \node[block, below of=p1, node distance=.6cm] (p2)  {$\inc{(\lift{\pi})}$};
  \node[block, right of=p2] (s2) {$\inc{(\lift{\sigma})}$};
  \node[below of=pj, node distance=.6cm] (mid) {};
  \node[block, circle, right of=mid, inner sep=0cm, node distance=1.2cm] (plus) {$+$};
  \node[block, right of=plus] (d) {$\inc{(\lift{\distinct})}$};
  \draw[->] (E) -- (j);
  \draw[->] (j) -- (pj);
  \draw[->] (E) -- (p1);
  \draw[->] (p1) -- (s1);
  \draw[->] (E) -- (p2);
  \draw[->] (p2) -- (s2);
  \draw[->] (E) -- (plus);
  \draw[->] (pj) -- (plus);
  \draw[->] (s1) -- (plus);
  \draw[->] (s2) -- (plus);
  \draw[->] (plus) -- (d);
  
  \node[block, right of=d] (S) {$\int$};
  \node[right of=S, node distance=.8cm] (output)  {\code{R}};
  \draw[->] (Einput) -- (E);
  \draw[->] (d) -- node (o) {} (S);
  \draw[->] (S) -- (output);
  \node[block, above of=j, node distance=.8cm] (z) {$\zm$};
  \draw[->] (o.center) |- (z);
  \draw[->] (z) -- (j);
\end{tikzpicture}

Using the linearity of $\lift\pi$ and $\lift\sigma$, this can be rewritten as an equivalent
circuit:

\noindent
\begin{tikzpicture}[>=latex, node distance=1.3cm]
  \node[] (Einput) {\code{E}};
  \node[block, right of=Einput, node distance=.8cm] (E) {$\delta_0$};
  
  \node[right of=E] (empty) {};
  \node[block, above of=empty, node distance=.6cm] (j) {$\inc{(\lift{\bowtie})}$};
  \node[block, right of=j] (pj) {$\lift{\pi}$};
  \node[block, below of=j, node distance=1cm] (p1) {$\lift{\pi}$};
  \node[block, right of=p1] (s1) {$\lift{\sigma}$};
  \node[block, below of=p1, node distance=.6cm] (p2)  {$\lift{\pi}$};
  \node[block, right of=p2] (s2) {$\lift{\sigma}$};
  \node[below of=pj, node distance=.6cm] (mid) {};
  \node[block, circle, right of=mid, inner sep=0cm, node distance=1.2cm] (plus) {$+$};
  \node[block, right of=plus] (d) {$\inc{(\lift{\distinct})}$};
  \draw[->] (E) -- (j);
  \draw[->] (j) -- (pj);
  \draw[->] (E) -- (p1);
  \draw[->] (p1) -- (s1);
  \draw[->] (E) -- (p2);
  \draw[->] (p2) -- (s2);
  \draw[->] (E) -- (plus);
  \draw[->] (pj) -- (plus);
  \draw[->] (s1) -- (plus);
  \draw[->] (s2) -- (plus);
  \draw[->] (plus) -- (d);
  
  \node[block, right of=d] (S) {$\int$};
  \node[right of=S, node distance=.8cm] (output)  {\code{R}};
  \draw[->] (Einput) -- (E);
  \draw[->] (d) -- node (o) {} (S);
  \draw[->] (S) -- (output);
  \node[block, above of=j, node distance=.8cm] (z) {$\zm$};
  \draw[->] (o.center) |- (z);
  \draw[->] (z) -- (j);
\end{tikzpicture}

To make this circuit into a streaming computation that evaluates a new transitive
closure for a stream of inputs \code{E}, we lift it entirely,
using Proposition~\ref{prop-lift-cycle}:

\noindent
\begin{tikzpicture}[>=latex, node distance=1.3cm]
  \node[] (Einput) {\code{E}};
  \node[block, right of=Einput, node distance=.8cm] (E) {$\lift{\delta_0}$};
  
  \node[right of=E] (empty) {};
  \node[block, above of=empty, node distance=.6cm] (j) {$\lift{\inc{(\lift{\bowtie})}}$};
  \node[block, right of=j] (pj) {$\lift{\lift{\pi}}$};
  \node[block, below of=j, node distance=1cm] (p1) {$\lift{\lift{\pi}}$};
  \node[block, right of=p1] (s1) {$\lift{\lift{\sigma}}$};
  \node[block, below of=p1, node distance=.6cm] (p2)  {$\lift{\lift{\pi}}$};
  \node[block, right of=p2] (s2) {$\lift{\lift{\sigma}}$};
  \node[below of=pj, node distance=.6cm] (mid) {};
  \node[block, circle, right of=mid, inner sep=0cm, node distance=1.2cm] (plus) {$+$};
  \node[block, right of=plus] (d) {$\lift{\inc{(\lift{\distinct})}}$};
  \draw[->] (E) -- (j);
  \draw[->] (j) -- (pj);
  \draw[->] (E) -- (p1);
  \draw[->] (p1) -- (s1);
  \draw[->] (E) -- (p2);
  \draw[->] (p2) -- (s2);
  \draw[->] (E) -- (plus);
  \draw[->] (pj) -- (plus);
  \draw[->] (s1) -- (plus);
  \draw[->] (s2) -- (plus);
  \draw[->] (plus) -- (d);
  
  \node[block, right of=d, node distance=1.5cm] (S) {$\lift{\int}$};
  \node[right of=S, node distance=.8cm] (output)  {\code{R}};
  \draw[->] (Einput) -- (E);
  \draw[->] (d) -- node (o) {} (S);
  \draw[->] (S) -- (output);
  \node[block, above of=j, node distance=.8cm] (z) {$\lift{\zm}$};
  \draw[->] (o.center) |- (z);
  \draw[->] (z) -- (j);
\end{tikzpicture}

We convert this circuit into an incremental circuit, which receives
in each transaction the changes to relation \code{E} and produces the 
corresponding changes to relation \code{R}:

\noindent
\begin{tikzpicture}[>=latex, node distance=1.3cm]
  \node[] (DE) {$\Delta$\code{E}};
  \node[block, right of=DE, node distance=.7cm] (Einput) {$\I$};
  \draw[->] (DE) -- (Einput);
  \node[block, right of=Einput, node distance=.8cm] (E) {$\lift{\delta_0}$};
  
  \node[right of=E, node distance=1.2cm] (empty) {};
  \node[block, above of=empty, node distance=.6cm] (j) {$\lift{\inc{(\lift{\bowtie})}}$};
  \node[block, right of=j] (pj) {$\lift{\lift{\pi}}$};
  \node[block, below of=j, node distance=1cm] (p1) {$\lift{\lift{\pi}}$};
  \node[block, right of=p1] (s1) {$\lift{\lift{\sigma}}$};
  \node[block, below of=p1, node distance=.6cm] (p2)  {$\lift{\lift{\pi}}$};
  \node[block, right of=p2] (s2) {$\lift{\lift{\sigma}}$};
  \node[below of=pj, node distance=.6cm] (mid) {};
  \node[block, circle, right of=mid, inner sep=0cm, node distance=1cm] (plus) {$+$};
  \node[block, right of=plus, node distance=1.2cm] (d) {$\lift{\inc{(\lift{\distinct})}}$};
  \draw[->] (E) -- (j);
  \draw[->] (j) -- (pj);
  \draw[->] (E) -- (p1);
  \draw[->] (p1) -- (s1);
  \draw[->] (E) -- (p2);
  \draw[->] (p2) -- (s2);
  \draw[->] (E) -- (plus);
  \draw[->] (pj) -- (plus);
  \draw[->] (s1) -- (plus);
  \draw[->] (s2) -- (plus);
  \draw[->] (plus) -- (d);
  
  \node[block, right of=d] (S) {$\lift{\int}$};
  \node[block, right of=S, node distance=.7cm] (OD) {$\D$};
  \node[right of=OD, node distance=.8cm] (output)  {$\Delta$\code{R}};
  \draw[->] (Einput) -- (E);
  \draw[->] (d) -- node (o) {} (S);
  \draw[->] (S) -- (OD);
  \draw[->] (OD) -- (output);
  \node[block, above of=j, node distance=.8cm] (z) {$\lift{\zm}$};
  \draw[->] (o.center) |- (z);
  \draw[->] (z) -- (j);
\end{tikzpicture}

We can now apply again the chain rule to this circuit:

\noindent
\begin{tikzpicture}[>=latex, node distance=1.4cm]
  \node[] (Einput) {$\Delta$\code{E}};
  \node[block, right of=Einput, node distance=1cm] (E) {$\inc{(\lift{\delta_0})}$};
  
  \node[right of=E] (empty) {};
  \node[block, above of=empty, node distance=.6cm] (j) {$\inc{(\lift{\inc{(\lift{\bowtie})}})}$};
  \node[block, right of=j, node distance=1.6cm] (pj) {$\inc{(\lift{\lift{\pi}})}$};
  \node[block, below of=j, node distance=1cm] (p1) {$\inc{(\lift{\lift{\pi}})}$};
  \node[block, right of=p1] (s1) {$\inc{(\lift{\lift{\sigma}})}$};
  \node[block, below of=p1, node distance=.6cm] (p2)  {$\inc{(\lift{\lift{\pi}})}$};
  \node[block, right of=p2] (s2) {$\inc{(\lift{\lift{\sigma}})}$};
  \node[below of=pj, node distance=.6cm] (mid) {};
  \node[block, circle, right of=mid, inner sep=0cm, node distance=1cm] (plus) {$+$};
  \node[block, right of=plus] (d) {$\inc{(\lift{\inc{(\lift{\distinct})}})}$};
  \draw[->] (E) -- (j);
  \draw[->] (j) -- (pj);
  \draw[->] (E) -- (p1);
  \draw[->] (p1) -- (s1);
  \draw[->] (E) -- (p2);
  \draw[->] (p2) -- (s2);
  \draw[->] (E) -- (plus);
  \draw[->] (pj) -- (plus);
  \draw[->] (s1) -- (plus);
  \draw[->] (s2) -- (plus);
  \draw[->] (plus) -- (d);
  
  \node[block, right of=d, node distance=1.8cm] (S) {$\inc{(\lift{\int})}$};
  \node[right of=S, node distance=1cm] (output)  {$\Delta$\code{R}};
  \draw[->] (Einput) -- (E);
  \draw[->] (d) -- node (o) {} (S);
  \draw[->] (S) -- (output);
  \node[block, above of=j, node distance=.8cm] (z) {$\inc{(\lift{\zm})}$};
  \draw[->] (o.center) |- (z);
  \draw[->] (z) -- (j);
\end{tikzpicture}

We now take advantage of the linearity of $\lift\delta_0$, $\lift\int$, 
$\lift\zm$, $\lift\lift\pi$, and $\lift\lift\sigma$ to simplify the circuit
by removing some $\inc{\cdot}$ invocations:

\noindent
\begin{tikzpicture}[>=latex, node distance=1.3cm]
  \node[] (Einput) {$\Delta$\code{E}};
  \node[block, right of=Einput, node distance=.8cm] (E) {$\lift{\delta_0}$};
  
  \node[right of=E] (empty) {};
  \node[block, above of=empty, node distance=.6cm] (j) {$\inc{(\lift{\inc{(\lift{\bowtie})}})}$};
  \node[block, right of=j, node distance=1.6cm] (pj) {$\lift{\lift{\pi}}$};
  \node[block, below of=j, node distance=1cm] (p1) {$\lift{\lift{\pi}}$};
  \node[block, right of=p1] (s1) {$\lift{\lift{\sigma}}$};
  \node[block, below of=p1, node distance=.6cm] (p2)  {$\lift{\lift{\pi}}$};
  \node[block, right of=p2] (s2) {$\lift{\lift{\sigma}}$};
  \node[below of=pj, node distance=.6cm] (mid) {};
  \node[block, circle, right of=mid, inner sep=0cm, node distance=.8cm] (plus) {$+$};
  \node[block, right of=plus] (d) {$\inc{(\lift{\inc{(\lift{\distinct})}})}$};
  \draw[->] (E) -- (j);
  \draw[->] (j) -- (pj);
  \draw[->] (E) -- (p1);
  \draw[->] (p1) -- (s1);
  \draw[->] (E) -- (p2);
  \draw[->] (p2) -- (s2);
  \draw[->] (E) -- (plus);
  \draw[->] (pj) -- (plus);
  \draw[->] (s1) -- (plus);
  \draw[->] (s2) -- (plus);
  \draw[->] (plus) -- (d);
  
  \node[block, right of=d, node distance=1.8cm] (S) {$\lift{\int}$};
  \node[right of=S, node distance=.8cm] (output)  {$\Delta$\code{R}};
  \draw[->] (Einput) -- (E);
  \draw[->] (d) -- node (o) {} (S);
  \draw[->] (S) -- (output);
  \node[block, above of=j, node distance=.8cm] (z) {$\lift{\zm}$};
  \draw[->] (o.center) |- (z);
  \draw[->] (z) -- (j);
\end{tikzpicture}

There are two applications of $\inc{\cdot}$ left in this circuit: $\inc{(\lift{\inc{(\lift{\bowtie})}})}$
and $\inc{(\lift{\inc{(\lift\distinct)}})}$.  We expand their implementations separately,
and we stitch them into the global circuit at the end.  This ability to reason about
sub-circuits independently highlights the modularity of \dbsp.

The join is expanded twice, using the bilinearity
of $\lift\bowtie$ and $\lift\lift\bowtie$.  Let's start with the inner circuit,
implementing $\inc{(\lift{\bowtie})}$, given by Theorem~\ref{bilinear}:

\begin{tabular}{m{2cm}m{.5cm}m{4.5cm}}
\begin{tikzpicture}[auto,>=latex]
    \node[] (a) {$a$};
    \node[below of=a, node distance=.6cm] (midway) {};
    \node[below of=midway, node distance=.6cm] (b) {$b$};
    \node[block, right of=midway] (q) {$\inc{(\lift{\bowtie})}$};
    \node[right of=q] (output) {$o$};
    \draw[->] (a) -| (q);
    \draw[->] (b) -| (q);
    \draw[->] (q) -- (output);
\end{tikzpicture} &
$\cong$ &
\begin{tikzpicture}[auto,>=latex]
  \node[] (a) {$a$}; 
  \node[block, below of=a, node distance=.6cm] (ab) {$\lift\bowtie$};
  \node[below of=ab, node distance=.6cm] (b) {$b$};
  \node[block, right of=a, node distance=1cm] (jI1) {$\I$};
  \node[block, right of=b, node distance=1cm] (jI2) {$\I$};
  \draw[->] (a) -- (jI1);
  \draw[->] (b) -- (jI2);
  \node[block, right of=jI1, node distance=.8cm] (ZI1) {$\zm$};
  \node[block, right of=jI2, node distance=.8cm] (ZI2) {$\zm$};
  \draw[->] (jI1) -- (ZI1);
  \draw[->] (jI2) -- (ZI2);
  \node[block, right of=ZI1] (DI1) {$\lift\bowtie$};
  \node[block, right of=ZI2] (DI2) {$\lift\bowtie$};
  \draw[->] (ZI1) -- (DI1);
  \draw[->] (ZI2) -- (DI2);
  \node[block, circle, below of=DI1, inner sep=0cm, node distance=.6cm] (sum) {$+$};
  \node[right of=sum] (output) {$o$};
  \draw[->] (ab) -- (sum);
  \draw[->] (DI1) -- (sum);
  \draw[->] (DI2) -- (sum);
  \draw[->] (a) -- (ab);
  \draw[->] (b) -- (ab);
  \draw[->] (a) -- (DI2);
  \draw[->] (b) -- (DI1);
  \draw[->] (sum) -- (output);
\end{tikzpicture}
\end{tabular}

Now we lift and incrementalize to get the circuit for $\inc{(\lift{\inc{(\lift{\bowtie})}})}$:

\begin{tikzpicture}[auto,>=latex]
  \node[] (a) {$a$}; 
  \node[below of=a, node distance=.8cm] (midway) {};
  \node[below of=ab, node distance=.8cm] (b) {$b$};

  \node[block, right of=a] (Ia) {$\I$};
  \node[block, right of=b] (Ib) {$\I$};
  \draw[->] (a) -- (Ia);
  \draw[->] (b) -- (Ib);
    
  \node[block, below of=Ia, node distance=.7cm] (ab) {$\lift\lift\bowtie$};
  \node[block, right of=Ia, node distance=1cm] (jI1) {$\lift\I$};
  \node[block, right of=Ib, node distance=1cm] (jI2) {$\lift\I$};
  \draw[->] (Ia) -- (jI1);
  \draw[->] (Ib) -- (jI2);
  \node[block, right of=jI1] (ZI1) {$\lift\zm$};
  \node[block, right of=jI2] (ZI2) {$\lift\zm$};
  \draw[->] (jI1) -- (ZI1);
  \draw[->] (jI2) -- (ZI2);
  \node[block, right of=ZI1] (DI1) {$\lift\lift\bowtie$};
  \node[block, right of=ZI2] (DI2) {$\lift\lift\bowtie$};
  \draw[->] (ZI1) -- (DI1);
  \draw[->] (ZI2) -- (DI2);
  \node[block, circle, below of=DI1, inner sep=0cm, node distance=.7cm] (sum) {$+$};
  \node[block, right of=sum] (D) {$\D$};
  \node[right of=D] (output) {$o$};
  \draw[->] (ab) -- (sum);
  \draw[->] (DI1) -- (sum);
  \draw[->] (DI2) -- (sum);
  \draw[->] (Ia) -- (ab);
  \draw[->] (Ib) -- (ab);
  \draw[->] (Ia) -- (DI2);
  \draw[->] (Ib) -- (DI1);
  \draw[->] (sum) -- (D);
  \draw[->] (D) -- (output);
\end{tikzpicture}

Applying the chain rule and the linearity of $\lift\I$ and $\lift\zm$ this becomes:

\begin{tikzpicture}[auto,>=latex, node distance=1.2cm]
  \node[] (a) {$a$}; 
  \node[block, below of=a, node distance=.8cm] (ab) {$\inc{(\lift\lift\bowtie)}$};
  \node[below of=ab, node distance=.8cm] (b) {$b$};
  \node[block, right of=a, node distance=1cm] (jI1) {$\lift{\I}$};
  \node[block, right of=b, node distance=1cm] (jI2) {$\lift{\I}$};
  \draw[->] (a) -- (jI1);
  \draw[->] (b) -- (jI2);
  \node[block, right of=jI1] (ZI1) {$\lift{\zm}$};
  \node[block, right of=jI2] (ZI2) {$\lift{\zm}$};
  \draw[->] (jI1) -- (ZI1);
  \draw[->] (jI2) -- (ZI2);
  \node[block, right of=ZI1] (DI1) {$\inc{(\lift\lift\bowtie)}$};
  \node[block, right of=ZI2] (DI2) {$\inc{(\lift\lift\bowtie)}$};
  \draw[->] (ZI1) -- (DI1);
  \draw[->] (ZI2) -- (DI2);
  \node[block, circle, below of=DI1, inner sep=0cm, node distance=.8cm] (sum) {$+$};
  \node[right of=sum] (output) {$o$};
  \draw[->] (ab) -- (sum);
  \draw[->] (DI1) -- (sum);
  \draw[->] (DI2) -- (sum);
  \draw[->] (a) -- (ab);
  \draw[->] (b) -- (ab);
  \draw[->] (a) -- (DI2);
  \draw[->] (b) -- (DI1);
  \draw[->] (sum) -- (output);
\end{tikzpicture}

We now have three applications of $\inc{(\lift\lift\bowtie)}$.  Each of these is the
incremental form of a bilinear operator, so it looks like in the end we will have $3\times3$ 
applications of $\lift\lift\bowtie$.  In fact, the overall expression can be simplified
(see~\cite{tr} for a precise derivation), and the end result only has 4 terms in $\lift\lift\bowtie$.

Here is the final form of the expanded join circuit:

\begin{tikzpicture}[auto,>=latex]
  \node[] (a) {$a$}; 
  \node[below of=a, node distance=.8cm] (b) {$b$};
  
  \node[block, right of=a] (LIa) {$\lift{\I}$};
  \node[block, above of=LIa, node distance=.8cm] (Ia) {$\I$};
  \node[block, right of=LIa] (IIa) {$\I$};
  \node[block, right of=Ia] (zIa) {$\zm$};
  \draw[->] (a) -- (LIa);
  \draw[->] (a) -- (Ia);
  \draw[->] (Ia) -- (zIa);
  \draw[->] (LIa) -- (IIa);
  
  \node[block, right of=b] (Ib) {$\I$};
  \node[block, below of=Ib, node distance=.8cm] (LIb) {$\lift\I$};
  \node[block, right of=Ib] (zb) {$\zm$};
  \node[block, right of=LIb] (IIb) {$\I$};
  \node[block, right of=IIb] (zIIb) {$\lift\zm$};
  \node[block, below of=IIb] (zIb) {$\lift\zm$};
  \draw[->] (b) -- (Ib);
  \draw[->] (b) -- (LIb);
  \draw[->] (Ib) -- (zb);
  \draw[->] (LIb) -- (IIb);
  \draw[->] (IIb) -- (zIIb);
  \draw[->] (LIb) -- (zIb);
  
  \node[block, right of=zIIb] (j1) {$\lift\lift\bowtie$};
  \node[block, above of=j1, node distance=.8cm]   (j2) {$\lift\lift\bowtie$};
  \node[block, above of=j2, node distance=.8cm]   (j3) {$\lift\lift\bowtie$};
  \node[block, above of=j3, node distance=.8cm]   (j4) {$\lift\lift\bowtie$};
  \draw[->] (zIIb) -- (j1);
  \draw[->] (a) -- (j1);
  \draw[->] (zb) -- (j2);
  \draw[->] (LIa) -- (j2);
  \draw[->] (IIa) -- (j3);
  \draw[->] (b) -- (j3);
  \draw[->] (zIa) -- (j4);
  \draw[->] (zIb) -- (j4);
  
  \node[block, right of=j3] (plus) {$+$};
  \draw[->] (j1) -- (plus);
  \draw[->] (j2) -- (plus);
  \draw[->] (j3) -- (plus);
  \draw[->] (j4) -- (plus); 
  \node[right of=plus] (o) {$o$};
  \draw[->] (plus) -- (o);
\end{tikzpicture}

Returning to $\inc{(\lift{\inc{(\lift\distinct)}})}$, we can compute its circuit by expanding
once using Proposition~\ref{prop-inc_distinct}:

\noindent
\begin{tabular}{m{3cm}m{.5cm}m{3.5cm}}
\begin{tikzpicture}[>=latex]
\node[] (input) {$i$};
\node[block, right of=input, node distance=1.5cm] (d) {$\inc{(\lift{\inc{(\lift{\distinct})}})}$};
\node[right of=d, node distance=1.5cm] (output) {$o$};
\draw[->] (input) -- (d);
\draw[->] (d) -- (output);
\end{tikzpicture}
& $\cong$ &
\begin{tikzpicture}[>=latex, node distance=.8cm]
    \node[] (input) {$i$};
    \node[block, right of=input] (I) {$\I$};
    \node[block, right of=I] (LI) {$\lift{\I}$};
    \node[block, right of=LI, node distance=1cm] (z) {$\lift{\zm}$};
    \node[block, below of=z, node distance=.8cm] (H) {$\lift{\lift{H}}$};
    \node[block, right of=H] (D) {$\D$};
    \node[right of=D] (output) {$o$};
    \draw[->] (input) -- (I);
    \draw[->] (I) -- node (mid) {} (LI);
    \draw[->] (LI) -- (z);
    \draw[->] (mid.center) |- (H);
    \draw[->] (z) -- (H);
    \draw[->] (H) -- (D);
    \draw[->] (D) -- (output);
\end{tikzpicture}
\end{tabular}

Finally, stitching all these pieces together we get the final circuit
shown in Figure~\ref{fig:recursive-example}.

\begin{figure*}[h]
\begin{tikzpicture}[>=latex]
  \node[] (Einput) {$\Delta$\code{E}};
  \node[block, right of=Einput, node distance=.8cm] (dE) {$\lift{\delta_0}$};
  \draw[->] (Einput) -- (dE); 
  \node[right of=dE] (empty) {};

  \node[block, above of=empty, node distance=2.5cm] (b) { };
  \node[block, above of=b, node distance=.8cm] (a) { };
  \draw[->] (dE) -- (b);
  
  \node[block, right of=a] (LIa) {$\lift{\I}$};
  \node[block, above of=LIa, node distance=.8cm] (Ia) {$\I$};
  \node[block, right of=LIa] (IIa) {$\I$};
  \node[block, right of=Ia] (zIa) {$\zm$};
  \draw[->] (a) -- (LIa);
  \draw[->] (a) -- (Ia);
  \draw[->] (Ia) -- (zIa);
  \draw[->] (LIa) -- (IIa);
  
  \node[block, right of=b] (Ib) {$\I$};
  \node[block, below of=Ib, node distance=.8cm] (LIb) {$\lift\I$};
  \node[block, right of=Ib] (zb) {$\zm$};
  \node[block, right of=LIb] (IIb) {$\I$};
  \node[block, right of=IIb] (zIIb) {$\lift\zm$};
  \node[block, below of=IIb] (zIb) {$\lift\zm$};
  \draw[->] (b) -- (Ib);
  \draw[->] (b) -- (LIb);
  \draw[->] (Ib) -- (zb);
  \draw[->] (LIb) -- (IIb);
  \draw[->] (IIb) -- (zIIb);
  \draw[->] (LIb) -- (zIb);
  
  \node[block, right of=zIIb] (j1) {$\lift\lift\bowtie$};
  \node[block, above of=j1, node distance=.8cm]   (j2) {$\lift\lift\bowtie$};
  \node[block, above of=j2, node distance=.8cm]   (j3) {$\lift\lift\bowtie$};
  \node[block, above of=j3, node distance=.8cm]   (j4) {$\lift\lift\bowtie$};
  \draw[->] (zIIb) -- (j1);
  \draw[->] (a) -- (j1);
  \draw[->] (zb) -- (j2);
  \draw[->] (LIa) -- (j2);
  \draw[->] (IIa) -- (j3);
  \draw[->] (b) -- (j3);
  \draw[->] (zIa) -- (j4);
  \draw[->] (zIb) -- (j4);
  
  \node[block, right of=j3, circle, inner sep=0cm] (plus) {$+$};
  \draw[->] (j1) -- (plus);
  \draw[->] (j2) -- (plus);
  \draw[->] (j3) -- (plus);
  \draw[->] (j4) -- (plus); 
  
  \node[block, right of=plus] (pj) {$\lift{\lift{\pi}}$};
  \node[block, below of=empty] (p1) {$\lift{\lift{\pi}}$};
  \node[block, right of=p1] (s1) {$\lift{\lift{\sigma}}$};
  \node[block, below of=p1, node distance=.6cm] (p2)  {$\lift{\lift{\pi}}$};
  \node[block, right of=p2] (s2) {$\lift{\lift{\sigma}}$};
  \node[below of=pj, node distance=.6cm] (mid) {};
  \node[block, circle, right of=empty, inner sep=0cm, node distance=8cm] (relplus) {$+$};
  
  \draw[->] (plus) -- (pj);
  \draw[->] (dE) -- (p1);
  \draw[->] (p1) -- (s1);
  \draw[->] (dE) -- (p2);
  \draw[->] (p2) -- (s2);
  \draw[->] (dE) -- (relplus);
  \draw[->] (pj) -- (relplus);
  \draw[->] (s1) -- (relplus);
  \draw[->] (s2) -- (relplus);
  
    \node[block, right of=relplus] (distI) {$\I$};
    \node[block, right of=distI] (distLI) {$\lift{\I}$};
    \node[block, right of=distLI, node distance=1cm] (distz) {$\lift{\zm}$};
    \node[block, below of=distz, node distance=.8cm] (distH) {$\lift{\lift{H}}$};
    \node[block, right of=distH] (distD) {$\D$};
    \draw[->] (relplus) -- (distI);
    \draw[->] (distI) -- node (distmid) {} (distLI);
    \draw[->] (distLI) -- (distz);
    \draw[->] (distmid.center) |- (distH);
    \draw[->] (distz) -- (distH);
    \draw[->] (distH) -- (distD);
  
  \node[block, right of=distD] (S) {$\lift{\int}$};
  \node[right of=S, node distance=.8cm] (output)  {$\Delta$\code{R}};
  \draw[->] (distD) -- node (o) {} (S);
  \draw[->] (S) -- (output);
  \node[block, above of=a, node distance=1.2cm] (z) {$\lift{\zm}$};
  \draw[->] (o.center) |- (z);
  \draw[->] (z) -- (a);
\end{tikzpicture}
\caption{Final form of circuit from \refsec{sec:recursive-example}\label{fig:recursive-example}.}
\end{figure*}

\subsection{Complexity of incremental recursive queries}

\paragraph{Time complexity}

The time complexity of an incremental recursive query can be estimated as a product of
the number of fixed point iterations and the complexity of each iteration. The
incrementalized circuit (\ref{eq:increcursive}) performs the same number of
iterations as the non-incremental circuit (\ref{eq:seminaive}) in the worst case:
once the non-incremental circuit reaches the fixed point, its output is constant
and so is its derivative computed by the incrementalized circuit.

Consider a nested stream of changes $s \in \stream{\stream{A}}, s[t_1][t_2]$,
where $t_1$ is the input timestamp and $t_2$ is the fixed point iteration number.
The unoptimized loop body $\inc{(\lift{\inc{(\lift{T})}})} = 
\D \circ \lift{\D} \circ \lift{\lift{T}}
\circ \lift{\I} \circ \I$ has the same time complexity as $T$ applied to the
aggregated input of size $R(s)[t_1][t_2] \defn \norm{(\lift{\I} \circ
\I)(s)[t_1][t_2]} = \norm{\sum_{(i_1,i_2) \leq (t_1, t_2)} s[i_1][i_2]}$.  As
before, an optimized circuit can be significantly more efficient.  For instance,
by applying Theorem~\ref{bilinear} twice, to $\bowtie$ and $\lift{\bowtie}$, we
obtain a circuit for nested incremental join $s_1
\inc{(\lift{\inc{(\lift{\bowtie})}})} s_2$ that runs in
$O(\norm{\lift{\I}(s1)[t1][t2]} \times \norm{\I(s2)[t1][t2]}) \ll 
O(R(s_1) \times R(s_2))$ (because each term is correspondingly smaller).

\paragraph{Space complexity} Integration ($\I$) and differentiation ($\D$) of a
stream $s \in \stream{\stream{A}}$ uses memory proportional to
$\sum_{t_2}\norm{\sum_{t_1}s[t_1][t_2]}$, i.e., the total size of changes
aggregated over columns of the matrix.  The unoptimized circuit integrates
and differentiates respectively inputs and outputs of the recursive program
fragment.  As we move $\I$ and $\D$ inside the circuit using the chain rule, we
additionally store changes to intermediate streams.  Effectively we cache results of 
fixed point iterations from earlier timestamps to update them efficiently as new input changes arrive.
Notice that space is proportional to the number of iterations of the inner while loop.

%% file: extensions.tex
\section{Extensions}\label{sec:extensions}

The \dbsp language can express a richer class of streaming computations (both incremental and non-incremental) than those covered so far. In this section we give several examples.

\subsection{Multisets and bags}

In \secref{sec:relational} we have shown how to implement the relational algebra on sets.
Some SQL queries however produce \emph{multisets}, e.g., \code{UNION ALL}.
Since \zrs generalize multisets and bags, it is easy to implement query
operators that compute on such structures.  For example, while SQL \code{UNION}
is \zr addition followed by $\distinct$, \code{UNION ALL} is just \zr addition.

\subsection{Aggregation}\label{sec:aggregation}

Aggregation in SQL applies a function $a$ to a whole set producing a ``scalar''
result with some type $R$: $a: 2^A \to R$.  We convert such aggregation
functions to operate on \zrs, so in \dbsp an aggregation function has
a signature $a: \Z[A] \to R$.  Correctness of the implementation is 
defined as in \refsec{sec:correctness}.

The SQL \texttt{COUNT} aggregation function is implemented on \zrs by $a_\texttt{COUNT} : \Z[A] \to \Z$, which
computes a \emph{sum} of all the element weights: $a_\texttt{COUNT}(s) = \sum_{x \in s} s[x]$.
The SQL \texttt{SUM} aggregation function is implemented on \zrs by $a_\texttt{SUM}: \Z[\R] \to \R$ which
performs a \emph{weighted sum} of all (real) values: $a_\texttt{SUM}(s) = \sum_{x \in s} x \times s[x]$.

With this definition the aggregation functions $a_\texttt{COUNT}$ and $a_\texttt{SUM}$ are in
fact linear transformations between the group $\Z[A]$ and the result group ($\Z$, and $\R$ respectively).

If the output of the \dbsp circuit can be such a ``scalar'' value, then aggregation
with a linear function is simply function application, and thus it is automatically incremental.  However, in general, for composing multiple queries
we require the result of an aggregation to be a singleton \zr (containing a single value),
and not a scalar value.  In this case the aggregation function is implemented in
\dbsp as the composition of the actual aggregation and the 
$\makeset: A \to \Z[A]$ function, 
which converts a scalar value of type $A$ to a singleton \zr, defined as follows:
$\makeset(x) \defn 1 \cdot x$.

In conclusion, the following SQL query: 
\code{SELECT SUM(c) FROM I} 
is implemented as the following circuit:

\begin{tikzpicture}[auto,>=latex]
  \node[] (I) {\code{I}};
  \node[block, right of=I] (pi) {$\pi_\texttt{C}$};
  \node[block, right of=pi] (a) {$a_\texttt{SUM}$};
  \draw[->] (I) -- (pi);
  \draw[->] (pi) -- (a);
  \node[block, right of=a, node distance=1.5cm] (m) {$\makeset$};
  \node[right of=m] (O) {\code{O}};
  \draw[->] (a) -- (m);
  \draw[->] (m) -- (O); 
\end{tikzpicture}

The lifted incremental version of this circuit is interesting: since $\pi$ 
and $a_\texttt{SUM}$ are linear, they are equivalent to their own incremental 
versions.  Although $\inc{(\lift \makeset)} = \D \circ \lift{\makeset} \circ \I$
cannot be simplified, it is nevertheless efficient, doing only O(1) work per
invocation, since its input and output are singleton values.

An aggregation function such as \texttt{AVG} can be written as the composition of 
a more complex linear function that computes a pair of values using 
\texttt{SUM} and \texttt{COUNT}, followed by a $\mbox{makeset}$ and a selection operation
that divides the two columns.

\begin{lstlisting}[language=SQL]
SELECT AVG(c) FROM I 
\end{lstlisting}

\begin{tikzpicture}[auto,>=latex]
  \node[] (I) {\code{I}};
  \node[block, right of=I] (pi) {$\pi_\texttt{C}$};
  \node[block, right of=pi, node distance=1.4cm] (sc) {$(a_\texttt{SUM}, a_\texttt{COUNT})$};
  \draw[->] (I) -- (pi);
  \draw[->] (pi) -- (sc);
  \node[block, right of=sc, node distance=1.8cm] (m) {$\makeset$};
  \node[block, right of=m, node distance=1.2cm] (div) {$\sigma_/$};
  \node[right of=div] (O) {\code{O}};
  \draw[->] (sc) -- (m);
  \draw[->] (m) -- (div);
  \draw[->] (div) -- (O);
\end{tikzpicture}

Finally, some aggregate functions, such as \code{MIN}, are 
\emph{not} incremental in general, since for handling deletions
they may need to know the full set, and not just its changes.  The lifted
incremental version of such aggregate functions is implemented essentially
by ``brute force'', using the formula $\inc{(\lift a_\texttt{MIN})}
= \D \circ \lift{a_\texttt{MIN}} \circ \I$.  Such functions perform work
proportional to $R(s)$ at each invocation.

Note that the SQL \code{ORDER BY} directive can be modeled as
a non-linear aggregate function that emits a list.  However, such an implementation it is not efficiently incrementalizable in \dbsp. 
We leave the efficient handling of ORDER BY to future work.

Even when aggregation results do not form a group, they usually form
a structure with a zero element.  We expect that a well-defined
aggregation function maps empty \zrs to zeros in the target domain.

\subsection{Grouping; indexed relations}\label{sec:grouping}

Pick an arbitrary set $K$ of ``key values.''
Consider the mathematical structure of finite maps from $K$ 
to \zrs over some other domain $A$: $K \to \Z[A] = \Z[A][K]$.
We call values $i$ of this structure \defined{indexed \zrs}: for
each key $k \in K$, $i[k]$ is a \zr.  Because 
the codomain $\Z[A]$ is an abelian group, this structure is itself 
an abelian group.

We use this structure to model the SQL \texttt{GROUP BY} operator in \dbsp.  
Consider a \defined{partitioning function}
$p: A \to K$ that assigns a key to any value in $A$.  We define the grouping function
$G_p: \Z[A] \to (K \to \Z[A])$ as $G_p(a)[k] \defn \sum_{x \in a.p(x)=k}a[x] \cdot x$.
When applied to a \zr $a$ this function returns a indexed \zr, where each element 
is called a \defined{grouping}\footnote{We use
``group'' for the algebraic structure and ``grouping'' for the result of \code{GROUP BY}.}: for each key $k$ a 
grouping is a \zr containing all elements of $a$ that map to $k$ 
(as in SQL, groupings are multisets, represented by \zrs).
Consider our example \zr $R$ from \refsec{sec:relational},
and a key function $p(s)$ that returns the first letter of the string 
$s$. Then we have that $G_p(R) = \{ \code{j} \mapsto \{ \code{joe} 
\mapsto 1 \}, \code{a} \mapsto \{ \code{anne} \mapsto -1 \} \}$,
i.e., grouping with this key function produces an indexed \zr with two groupings, each 
of which contains a \zr with one element.

The grouping function $G_p$ is linear for any $p$.
It follows that the group-by implementation in DBSP is automatically
incremental: given some changes
to the input relation we can apply the partitioning function
to each change separately to compute how each grouping changes.

\subsection{\texttt{GROUP BY-AGGREGATE}}

Grouping in SQL is almost always followed by aggregation.  
Let us consider an aggregation function $a: (K \times \Z[A]) \to B$ that produces values
in some group $B$, and an indexed relation of type $\Z[A][K]$, as defined above in~\refsec{sec:grouping}.
The nested relation aggregation operator $Agg_a: \Z[A][K] \to B$ applies $a$ 
to the contents of each grouping independently and adds the results:
$Agg_a(g) \defn \sum_{k \in K} a(k, g[k])$.  To apply this
to our example, let us compute the equivalent of GROUP-BY count; we use
the following aggregation function $count: K \times \Z[A]$, $count(k, s) = 
\makeset((k, a_\texttt{COUNT}(s)))$, using the \zr counting function $a_\texttt{COUNT}$ 
from~\refsec{sec:aggregation}; the notation $(a,b)$ is a pair of values $a$ and $b$.
Then we have $Agg_{count} (G_p(R)) = \{ (\code{j}, 1) \mapsto 1, 
(\code{a}, -1) \mapsto 1 \}$.

Notice that, unlike SQL, \dbsp can express naturally computations
on indexed \zrs, they are just an instance of a group structure. 
One can even implement queries that operate on each grouping 
in an indexed \zr.  However, our definition of incremental 
computation is only concerned with incrementality in the 
\emph{outermost} structures.  We leave it to future work to
explore an appropriate definition of incremental computation that
operates on the \emph{inner} relations.

A very useful operation on nested relations is \defined{flatmap}, which is
essentially the inverse of partitioning, converting an indexed
\zr into a \zr: $\mbox{flatmap}: \Z[A][K] \to \Z[A \times K]$.
$\mbox{flatmap}$ is in fact a particular instance of aggregation,
using the aggregation function $a: K \times \Z[A] \to \Z[A \times K]$
defined by $a(k, s) = \sum_{x \in s[k]} s[k][x] \cdot (k, x).$
For our previous example, $\mbox{flatmap}(G_p(R)) = \{ (\code{j}, \code{joe}) \mapsto 1, 
(\code{a}, \code{anne}) \mapsto -1 \}$.

If we use an aggregation function $a: K \times Z[A]$ that is linear in its
second argument, then the aggregation operator $Agg_a$ is linear, and
thus fully incremental.  As a consequence, $\mbox{flatmap}$ is linear.  
However, many practical aggregation functions for nested relations are in fact 
not linear; an example is the $count$ function above, which is not linear
since it uses the $\makeset$ non-linear function.  Nevertheless, while 
the incremental evaluation of such functions is not fully incremental, 
it is at least partly incremental: when applying a change to groupings, the aggregation 
function only needs to be re-evaluated \emph{for groupings that have changed}.

\subsection{Antijoin}\label{sec:antijoin}\index{antijoin}

Antijoins arise in the implementation of Datalog programs with stratified negation.
Consider the following program:

\begin{lstlisting}[language=ddlog]
O(v, z) :- I1(v, z), not I2(v).     
\end{lstlisting}

The semantics of such a rule is defined in terms of joins and set difference.
This rule is equivalent with the following pair of rules:

\begin{lstlisting}[language=ddlog]
C(v, z) :- I1(v, z), I2(v).
O(v, z) :- I1(v, z), not C(v, z).     
\end{lstlisting}

This transformation reduces an antijoin to a join 
followed by a set difference.  This produces the following \dbsp circuit:

\begin{tikzpicture}[auto,>=latex]
  \node[] (i1) {\code{I1}};
  \node[below of=i1, node distance=.5cm] (i2) {\code{I2}};
  \node[block, right of=i1, node distance=1.5cm] (join) {$\bowtie$};
  \node[block, shape=circle, inner sep=0in, right of=join] (m) {---};
  \node[block, above of=m, shape=circle, inner sep=0in, node distance=.6cm] (plus) {$+$};
  \node[block, right of=plus, node distance=1cm] (distinct) {$\distinct$};
  \node[right of=distinct, node distance=1cm] (output) {\code{O}};
  \draw[->] (i1) -- node (tap) {} (join);
  \draw[->] (i2) -| (join);
  \draw[->] (join) -- (m);
  \draw[->] (m) -- (plus);
  \draw[->] (tap.south) |- (plus);
  \draw[->] (plus) -- (distinct);
  \draw[->] (distinct) -- (output);
\end{tikzpicture}

\subsection{Streaming joins}

Consider a binary query $T(s, t) = \I(s)~~\lift{\bowtie}~~t$.  This is the
\emph{relation-to-stream join} operator supported by streaming databases like ksqlDB~\cite{jafarpour-edbt19}.
Stream $s$ carries changes to a relation, while $t$ carries arbitrary data, e.g., logs
or telemetry data points. $T$ discards values from $t$ after matching them against the accumulated contents of the relation.


\subsubsection{Streaming Window queries}

Streaming databases often organize the contents of streams into windows, which store a subset of data points with a predefined range of timestamps.
The circuit below (a convolution filter in DSP) computes a \emph{fixed-size sliding-window aggregate}
over the last four timestamps defined by the $T_i$ functions.

\begin{center}
\begin{tikzpicture}[>=latex]
    \node[] (input) {$s$};
    \node[block, right of=input, node distance=1.5cm] (f0) {$T_0$};
    \node[below of=input, node distance=.5cm] (fake) {};
    \node[block, right of=fake, node distance=1cm] (z0) {$\zm$};
    \node[right of=input, node distance=.35cm] (tap) {};
    \node[block, right of=f0, node distance=1.5cm] (f1) {$T_1$};
    \node[block, right of=z0, node distance=1.2cm] (z1) {$\zm$};
    \node[block, right of=f1, node distance=1.5cm] (f2) {$T_2$};
    \node[block, right of=z1, node distance=1.5cm] (z2) {$\zm$};
    \draw[->] (input) -- (f0);
    \draw[->] (tap.center) |- (z0);
    \draw[->] (z0) -| (f0);
    \draw[->] (f0) -- (f1);
    \draw[->] (z0) -- (z1);
    \draw[->] (z1) -| (f1);
    \draw[->] (f1) -- (f2);
    \draw[->] (z1) -- (z2);
    \draw[->] (z2) -| (f2);
    \node[right of=f2] (output) {$o$};
    \draw[->] (f2) -- (output);
\end{tikzpicture}
\end{center}

In practice, windowing is usually based on physical timestamps attached to
stream values rather than logical time.  For instance, the CQL~\cite{arasu-tr02} query
``\texttt{SELECT * FROM events [RANGE 1 hour]}'' returns all events received
within the last hour.  The corresponding circuit (on the left) takes input stream $s \in \stream{\Z[A]}$ and an additional
input $\theta \in \stream{\mathbb{R}}$ that carries the value of the current
time.

\begin{tabular}{m{3cm}m{0.5cm}m{3cm}}
\begin{tikzpicture}[>=latex]
    \node[] (input) {$s$};
    \node[above of=input, node distance=.5cm] (t) {$\theta$};
    \node[block, right of=input] (i) {$I$};
    \node[block, right of=i] (w) {$W$};
    \node[right of=w] (output) {$o$};
    \draw[->] (input) -- (i);
    \draw[->] (i) -- (w);
    \draw[->] (w) -- (output);
    \draw[->] (t) -| (w);
\end{tikzpicture}
&
$\cong$
&
\begin{tikzpicture}[>=latex]
    \node[] (input) {$s$};
    \node[above of=input, node distance=.5cm] (t) {$\theta$};
    \node[block, shape=circle, right of=input, inner sep=0pt] (plus) {$+$};
    \node[block, right of=plus] (w) {$W$};
    \node[right of=w] (output) {$o$};
    \node[block, below of=plus, node distance=.6cm] (z) {$\zm$};
    \draw[->] (input) -- (plus);
    \draw[->] (plus) -- (w);
    \draw[->] (t) -| (w);
    \draw[->] (w) -- node (mid) {} (output);
    \draw[->] (mid.center) |-  (z);
    \draw[->] (z) -- (plus);
\end{tikzpicture} \\
\end{tabular}

\noindent{}where the \emph{window operator} $W$ prunes input \zrs, only keeping values
with timestamps less than an hour behind $\theta[t]$.  Assuming $ts: A \to \mathbb{R}$ returns
the physical timestamp of a value, $W$ is defined as $W(v, \theta)[t] \defn \{x \in v[t] . 
ts(x) \geq \theta[t] - 1hr\}$.  Assuming $\theta$ increases monotonically, $W$
can be moved inside integration, resulting in the circuit on the right, which uses
bounded memory to compute a window of an unbounded stream.
This circuit is a building block of a large family of window queries, including
window joins and aggregation.  We conjecture that \dbsp can express 
any CQL query.

\subsection{Relational while queries}

(See also non-monotonic semantics for Datalog$^\neg$ and Datalog$^{\neg\neg}$\cite{Abiteboul-book95}.)
To illustrate the power of \dbsp we implement the following
``while'' program, where $Q$ is an arbitrary relational algebra query:
{\small
\begin{lstlisting}[language=Pascal]
x := i;
while (x changes)
    x := Q(x);
\end{lstlisting}}
The \dbsp implementation of this program is:

\begin{center}
\begin{tikzpicture}[>=latex]
  \node[] (input) {$i$};
  \node[block, right of=input] (delta) {$\delta_0$};
  \node[block, circle, right of=delta, inner sep=0cm] (p) {$+$};
  \node[block, right of=p] (Q) {$\lift Q$};
  \node[block, right of=Q] (D) {$\D$};
  \node[block, right of=D] (S) {$\int$};
  \node[right of=S] (output)  {$x$};
  \node[block, below of=p, node distance=.7cm] (z) {$\zm$};
  \draw[->] (input) -- (delta);
  \draw[->] (delta) -- (p);
  \draw[->] (p) -- (Q);
  \draw[->] (Q) -- node (mid) {} (D);
  \draw[->] (D) -- (S);
  \draw[->] (mid.center) |- (z);
  \draw[->] (S) -- (output);
  \draw[->] (z) -- (p);
\end{tikzpicture}
\end{center}

This circuit can be converted to a streaming circuit that computes a stream of values $i$ 
by lifting it; it can be incrementalized using Algorithm~\ref{algorithm-inc} to compute on changes of $i$:

\begin{center}
\begin{tikzpicture}[>=latex]
  \node[] (input) {$\Delta i$};
  \node[block, right of=input] (delta) {$\lift{\delta_0}$};
  \node[block, circle, right of=delta, inner sep=0cm] (p) {$+$};
  \node[block, right of=p] (Q) {$\inc{(\lift{\lift{Q}})}$};
  \node[block, right of=Q, node distance=1.5cm] (D) {$\lift{\D}$};
  \node[block, right of=D, node distance=1.1cm] (S) {$\lift{\int}$};
  \node[right of=S, node distance=1.2cm] (output)  {$\Delta x$};
  \node[block, below of=p, node distance=.7cm] (z) {$\lift{\zm}$};
  \draw[->] (input) -- (delta);
  \draw[->] (delta) -- (p);
  \draw[->] (p) -- (Q);
  \draw[->] (Q) -- node (mid) {} (D);
  \draw[->] (D) -- (S);
  \draw[->] (mid.center) |- (z);
  \draw[->] (S) -- (output);
  \draw[->] (z) -- (p);
\end{tikzpicture}
\end{center}

Note that at runtime the execution of this circuit is not guaranteed to terminate; 
however, if the circuit does terminate, it will produce the correct 
output, i.e., the least fixpoint of $Q$ that includes~$i$.

%% file: implementation.tex
\section{Implementation}\label{sec:implementation}

We are prototyping an implementation of \dbsp as part of an
open-source project with an MIT license: 
\url{https://github.com/vmware/database-stream-processor}.
The implementation is written in Rust.
The implementation consists of a library and a runtime.
The library provides APIs for basic algebraic data types:
such as groups, finite maps, \zr, indexed \zr.  
A separate circuit construction API allows users to 
create \dbsp circuits by placing operator nodes (corresponding to boxes in our diagrams)
and connecting them with streams, which correspond to the
arrows in our diagrams.  The library provides pre-built generic operators
for integration, differentiation, delay, nested integration and differentiation.
The \zr library provides functions for computing the basic \zr operations
corresponding to plus, negation, grouping, joining, aggregation.

For iterative computations we provide the $\delta_0$ operator and
an operator that approximates $\int$ by terminating iteration of
a loop at a user-specified condition (usually the condition is the 
requirement for a zero to appear in a specified stream).
The low level library allows users to construct incremental
circuits manually.  Our plan is to add a higher-level library
(a compiler) which will automatically incrementalize a circuit.

%% file: related.tex
\section{Related work}\label{sec:related}

\dbsp using non-nested streams is a simplified instance of a Kahn 
network~\cite{kahn-ifip74}.  Johnson~\cite{johnson-phd83}
studies a very similar computational model without nested streams and its 
expressiveness. The implementation of such streaming models of computation and their
relationship to dataflow machines has been studied by Lee~\cite{lee-ieee95}.
Lee~\cite{lee-ifip93} also introduced streams of streams and the $\lift{\zm}$ operator.

In \secref{sec:extensions} we have discussed the connections with window and stream database 
queries~\cite{arasu-tr02,aurora}.

Incremental view maintenance (e.g.~\cite{gupta-idb93}) is
surveyed in~\cite{gupta-idb95}; a large bibliography is present in~\cite{motik-ai19}. 
Its most formal aspect is propagating ``deltas'' through algebraic expressions:
$Q(R+\Delta R)=Q(R)+\Delta Q(R,\Delta R)$. This work eventually crystallized in~\cite{koch-pods16}. DBSP 
incrementalization is both more modular and more fine-grain since it deals with streams of updates. 
Both~\cite{koch-pods10} and~\cite{green-tcs11} use \zrs to uniformly model insertions/deletions.

Picallo et al.~\cite{picallo-scop19} provide a general solution to IVM for
rich languages.  \dbsp requires a group structure on the values operated on; 
this assumption has two major practical benefits: it simplifies the mathematics considerably
(e.g., Picallo uses monoid actions to model changes), and it provides a general, simple
algorithm (\ref{algorithm-inc}) for incrementalizing arbitrary programs.  The downside of 
\dbsp is that one has to find a suitable group structure (e.g., \zrs for sets) to ``embed'' 
the computation.  Picallo's notion of ``derivative'' is not unique: they need creativity to choose
the right derivative definition, we need creativity to find the right group structure.

Many heuristic algorithms were published for Datalog-like languages, e.g., 
counting based approaches~\cite{Dewan-iis92,motik-aaai15} that maintain the number of derivations,
DRed~\cite{gupta-sigmod93} and its variants~\cite{Ceri-VLDB91,Wolfson-sigmod91,%
Staudt-vldb96,Kotowski-rr11,Lu-sigmod95,Apt-sigmod87}, the backward-forward algorithm and variants~\cite{motik-aaai15,
Harrison-wdd92,motik-ai19}.  \dbsp is more general than these approaches.  
Interestingly, the \zrs multiplicities in our relational implementation 
are related to the counting-number-of-derivations approaches.

\dbsp is tightly related to Differential Dataflow (DD)~\cite{mcsherry-cidr13, murray-sosp13}
and its theoretical foundations~\cite{abadi-fossacs15} (and recently~\cite{mchserry-vldb20,chothia-vldb16}).
All \dbsp operators are based on DD operators.  DD's computational model is more powerful than
\dbsp, since it allows past values in a stream to be "updated".
In contrast, our model assumes that the inputs of a computation arrive in the time order while allowing 
for nested time domains via the modular lifting transformer.  However, \dbsp can express both
incremental and non-incremental computations; in essence \dbsp is ``deconstructing'' DD into 
simple component building blocks; the core Proposition~\ref{prop-inc-properties} and
the Algorithm based on it~\ref{algorithm-inc} are new contributions.

%% file: conclusions.tex
\section{Conclusions}\label{sec:conclusions}

In this paper we have introduced \dbsp, a model of computation based on 
infinite streams over abelian groups.  In this model streams are used 
to model consecutive snapshots of a database, consecutive changes (or transactions)
applied to a database, and consecutive values of loop-carried variables.

We have defined an abstract notion of incremental computation over streams, 
and defined the incrementalization operator
$\inc{\cdot}$, which transforms a stream computation $Q$ into its incremental version $\inc{Q}$.
The incrementalization operator has some very nice algebraic properties, which
can generate very efficient incremental implementations for linear and bilinear
computations.

We have then applied these tools to two domains: relational queries and recursive 
stratified queries.  This gave us a general algorithm for incrementalizing an arbitrary
query, including recursive queries.
However, we believe that both the incrementalization algorithm and \dbsp are even more
powerful and can apply to even richer classes of query languages, including 
languages operating on nested relations and streaming query languages.

%% file: extra.tex
\section{Supporting material}\label{sec:extra}

\subsection{Operations on nested streams}\label{sec-nested-examples}

\newcommand{\ssa}[1]{
\setsepchar{ }
\readlist\arg{#1}
\begin{bmatrix}
   \begin{array}{ccccccc}
        {[} & \arg[1] & \arg[2] & \arg[3] & \arg[4] & \cdots & {]} \\
        {[} & \arg[5] & \arg[6] & \arg[7] & \arg[8] & \cdots & {]} \\
        {[} & \arg[9] & \arg[10] & \arg[11] & \arg[12] & \cdots & {]} \\
        {[} & \arg[13] & \arg[14] & \arg[15] & \arg[16] & \cdots & {]} \\
        \multicolumn{7}{c}{\vdots}
   \end{array}
\end{bmatrix}
}

If a stream can be thought of as an infinite vector, a stream of streams can be thought of
as an ``matrix'' with an infinite number of rows, where each row is a stream.  
For example, we can depict the nested stream 
$i \in \stream{\stream{\N}}$ defined by $i[t_0][t_1] = t_0 + 2 t_1$ as:
$$ i = \ssa{0 1 2 3 2 3 4 5 4 5 6 7 6 7 8 9} $$

\noindent ($t_0$ is the column index, and $t_1$ is the row index).  Let us
perform some computations on nested streams to get used to them.  Lifting twice
a scalar function computes on elements of the matrix pointwise:

$$(\lift{\lift{(x \mapsto x \bmod 2)}})(i) = 
  \ssa{0 1 0 1 0 1 0 1 0 1 0 1 0 1 0 1}
$$

The $\I$ operator on $\stream{\stream{A}}$ is well-defined: it operates on rows
of the matrix, treating each row as a single value:

$$\I(i) = \ssa{0 1 2 3 2 4 6 8 6 9 12 15 12 16 20 24}$$

Lifting a stream operator computing on $\stream{A}$, 
such as $\I: \stream{A} \to \stream{A}$, also produces an operator on nested streams, but
this time computing on the columns of the matrix:
$\lift{\I}: \stream{\stream{A}} \to \stream{\stream{A}}.$

$$(\lift{\I})(i) = \ssa{0 1 3 6 2 5 9 14 4 9 15 22 6 13 21 30}$$

Similarly, we can apply $\D$ to nested streams $\D : \stream{\stream{A}} \to
\stream{\stream{A}}$, computing on rows of the matrix:

$$\D(i) = \ssa{0 1 2 3 2 2 2 2 2 2 2 2 2 2 2 2}$$

\noindent while $\lift{\D} : \stream{\stream{A}} \to \stream{\stream{A}}$
computes on the columns:

$$(\lift{\D})(i) = \ssa{0 1 1 1 2 1 1 1 4 1 1 1 6 1 1 1}$$

Similarly, $\zm$ and its lifted variant have different outcomes:

$$\zm(i) = \ssa{0 0 0 0 0 1 2 3 2 3 4 5 4 5 6 7}$$

Notice the following commutativity properties for integration and differentiation 
on nested streams: $\I \circ (\lift{\I}) = (\lift{\I}) \circ \I$ and 
$\D \circ (\lift{\D}) = (\lift{\D}) \circ \D$.

$$(\lift{\zm})(i) = \ssa{0 0 1 2 0 2 3 4 0 4 5 6 0 6 7 8}$$

$\zm$ commutes with $\lift{\zm}$:

$$(\lift{\zm})(\zm(i)) = \zm((\lift{\zm})(i)) = \ssa{0 0 0 0 0 0 1 2 0 2 3 4 0 4 5 6}$$


$$\D_{\stream{\stream{\N}}}(i) = (\D(\lift{\D}))(i) = \ssa{0 1 1 1 2 0 0 0 2 0 0 0 2 0 0 0}$$

$$\I_{\stream{\stream{\N}}}(i) = ((\lift{\I})(\I))(i)= \ssa{0 1 3 6 2 6 12 20 6 15 27 42 12 28 48 72}$$